\documentclass[letter,english]{article}
\usepackage{emulateapj,onecolfloat,graphics}

\begin{document}

\newcommand{\Dc}{\Delta_{\rm c}}

\twocolumn
[
\title{The mass function}
\author{Martin White}
\affil{Departments of Physics and Astronomy, University of California,
  Berkeley, CA 94720}

\begin{abstract}
\noindent
We present the mass functions for different mass estimators for a range of
cosmological models.  We pay particular attention to how universal the mass
function is, and how it depends on the cosmology, halo identification and
mass estimator chosen.  We investigate quantitatively how well we can relate
observed masses to theoretical mass functions.
\end{abstract}
 ]

\section{Introduction}

One of the most fundamental predictions of a theory of structure
formation is the number density of objects of a given mass, the mass
function.  Accurate mass functions are used in a number of areas in
cosmology; in studies of galaxy formation, in measures of volumes
(e.g.~galaxy lensing) and in attempts to infer the normalization of
the power spectrum, the statistics of the initial density field, the
density parameter or the equation-of-state of the dark energy from the
abundance of rich clusters.  One of the most intriguing aspects of the
mass function is that it appears universal, in suitably scaled units,
for a wide range of theories.
A complete understanding of this phenomenon currently eludes us.

If we are to attempt to use the mass function to infer cosmological parameters
from the abundance of objects of some given property, then we need to
understand how accurate our theory for the mass function is.
This involves understanding how to define the mass of an object in cosmology,
a task which is non-trivial as there is no clear boundary between a halo and
the surrounding large-scale structure in theories of hierarchical structure
formation.
The purpose of this paper is to calculate the mass functions for different
mass estimators for a range of cosmological models, to see how well these
statistics can be computed from semi-analytic theories and to investigate
quantitatively how to relate observed masses to theoretical mass functions.

We find that the mass function is only approximately `universal', and only
for a few mass estimators.
We discuss how accurately one can convert between different mass estimates,
so as to relate what can easily be measured to what can easily be predicted.
We also discuss the limitations which non-universality of the mass function
would place on cosmological parameter estimation were it not to be corrected
for.

The outline is as follows:
after a review of some background (\S\ref{sec:pressschechter})
we present mass functions, derived from N-body simulations
(\S\ref{sec:simulations}), for a variety of different mass definitions
(\S\ref{sec:massdef}).  We compare these mass definitions, based on mean
density contrast, with the concept of a virialized halo (\S\ref{sec:virial}).
We investigate how universal the mass function is (\S\ref{sec:universal})
for each different estimator and present fitting functions
(\S\ref{sec:fitting}) to the mass functions for 3 different cosmological
models.  We finish by considering the effect of clustering on the mass
function (\S\ref{sec:clustering}) and summarize our main results
(\S\ref{sec:conclusions}).

\section{Press-Schechter Theory} \label{sec:pressschechter}

We begin by reviewing the basic theory underlying the expectation that the
number density of halos, per unit comoving volume, should take a universal
form.  This expectation was first elucidated by Press \& Schechter~\cite{PS}
who combined the statistics of the initial density field with a model for
the evolution of perturbations based on spherical collapse of a top-hat
overdensity (see e.g.~Peacock \cite{Peacock} for a textbook treatment;
Bower \cite{Bow}; Peacock \& Heavens \cite{PeaHea}; Bond et al.~\cite{BCEK}
Lacey \& Cole \cite{LacCol1} for more details).
Specifically these authors advanced the {\it ansatz\/} that the fraction of
mass in halos more massive than $M$ is related to the fraction of the volume
in which the smoothed initial density field is above some threshold $\delta_c$.
A variety of smoothing windows and thresholds have been advocated, but the
most common is a top-hat window in real space and $\delta_c\simeq 1.69$.

The P-S mass function agrees relatively well with the results of numerical
simulations both for critical density models with power-law spectra and, more
surprisingly, for models without this self-similar evolution
(e.g.~Efstathiou et al.~\cite{EFWD}; Efstathiou \& Rees \cite{EfsRee};
White, Efstathiou \& Frenk \cite{WEF}; Lacey \& Cole~\cite{LacCol2};
Gelb \& Bertschinger \cite{GelBer}; Bond \& Myers \cite{BonMye}).
The P-S mass function and numerical results are known to deviate in detail at
both the high and low mass ends.
Refinements to this theory have been advanced, all of which relate the
abundance of collapsed objects to peaks in the initial density field in a
`universal' manner.
In the latest incarnation the mass function has been motivated by or
fit to large cosmological N-body simulations
(Sheth \& Tormen~\cite{SheTor}; Jenkins et al.~\cite{JFWCCECY};
hereafter JFWCCECY).

To fix our notation we recap briefly the ingredients in this model in the
next two sections.

\subsection{Top-hat collapse} \label{sec:tophat}

The spherical top-hat ansatz describes the formation of a collapsed object
by solving for the evolution of a sphere of uniform overdensity $\delta$ in
a smooth background of density $\bar{\rho}$.  By Birkhoff's theorem the
overdense region evolves as a positively curved Friedman universe whose
expansion rate is initially matched to that of the background.
The overdensity at first expands but, because it is overdense, the expansion
slows (relative to the background) and eventually halts before the region
begins to recollapse.
Technically the collapse proceeds to a singularity but it is assumed in a
``real'' object virialization occurs at twice\footnote{There is a small
correction to this in the presence of a cosmological constant which
contributes a $\Lambda r^2$ potential.}
the turn-around time, resulting in a sphere of half the turn-around radius.
In an Einstein-de Sitter model the overdensity (relative to the critical
density) at virialization is $\Delta_{\rm c}=18\pi^2\simeq 178$.  We shall
always use $\Delta_{\rm c}$ to indicate the overdensity relative to critical
of a virialized halo, which will be lower for smaller $\Omega_{\rm m}$.
A fitting function for $\Delta_{\rm c}$ for arbitrary $\Omega_{\rm m}$ and
$\Omega_\Lambda$ can be found in Pierpaoli, Scott \& White~\cite{PSW}.
Note that some authors use a different convention in which $\Delta_{\rm c}$
is specified relative to the background matter density -- our $\Delta_{\rm c}$
is $\Omega_{\rm m}$ times theirs and we shall come back to this point
in \S\ref{sec:massdef}.
The linear theory extrapolation of this overdensity is normally denoted
$\delta_{\rm c}$ and is $(3/20)(12\pi)^{2/3}\simeq 1.686$ in an
Einstein-de Sitter model.  This overdensity is often used as a threshold
parameter in PS theory and its extensions and has a very weak cosmology
dependence which is often neglected.
We shall return to some of these considerations in \S\ref{sec:virial}.

\subsection{Multiplicity function}

The mass function now comes from considering the statistics of the initial
density field and the top-hat model above.  Under the P-S {\it ansatz\/} all
of the cosmology dependence is contained within the rms density fluctuation,
$\sigma(M)$, smoothed with a top-hat filter on a scale\footnote{In
principle $R$ could be defined with respect to $\rho_{\rm crit}$, but this is
not the natural choice in the top-hat collapse model.}
$R^3=3M/4\pi \bar{\rho}$.  The multiplicity function,
\begin{equation}
  \nu\,f(\nu)d\nu \equiv {M\over 2\bar{\rho}} {dn\over dM} dM
  \qquad ,
\end{equation}
is a universal function of the peak height $\nu$ which is related
to the mass of the halo through
\begin{equation}
  \nu \equiv {\delta_c\over \sigma(M)}
\end{equation}
with $\delta_c=1.69$.
Note that some authors, particularly Sheth \& Tormen~\cite{SheTor}, define
$\nu$ to be $(\delta_c/\sigma(M))^2$ rather than $\delta_c/\sigma(M)$
as we have done.
If the initial fluctuations are Gaussian, as we shall assume throughout, then
the multiplicity function is simply
\begin{equation}
  \nu f(\nu) \propto e^{-\nu^2/2}
\end{equation}
where the normalization constant is fixed by the requirement that all of the
mass lie in a given halo
\begin{equation}
  \int \nu\, f(\nu) d\nu = {1\over 2} \qquad .
\end{equation}
There is no justification for this normalization from N-body simulations,
which cannot probe the $M\to 0$ tail, but we shall adopt it throughout.

Motivated by a model of elliptical collapse, Sheth \& Tormen~\cite{SheTor}
provided a fit to large, high-resolution N-body simulations of the
modified form
\begin{equation}
  f(\nu) \propto \left(1+(a\nu^2)^{-p}\right)(a\nu^2)^{-1/2}
    e^{-a \nu^2/2}
\label{eqn:fnu}
\end{equation}
where $p=0.3$ and $a=0.707$ provided the best fit to groups selected with a
spherical overdensity algorithm.
A slightly different fitting function was proposed by JFWCCECY based on
analysis of the same simulations.
The advantage of Eq.~(\ref{eqn:fnu}) over that of JFWCCECY is that it is well
behaved over the full range of mass, whereas the functional form of JFWCCECY
cannot be safely extrapolated outside of the range of their fit.
In addition the elliptical collapse model can be used to discuss the
clustering of halos (Sheth \& Tormen \cite{SheTor}) using an extension of
the peak-background split formalism
(e.g.~Efstathiou et al.~\cite{EFWD}; Cole \& Kaiser \cite{ColKai};
 Mo \& White \cite{MoWhi}).
While we shall see later that the Sheth \& Tormen form overpredicts the
number of small mass halos in some of our simulations, that region is not
the main focus of this work.
Small differences in the functional form of the fitting functions will not
be important for our conclusions.

\subsection{Toward higher accuracy}

To recap the material in the last 2 sub-sections, we assume that the mass
function of virialized objects depends only on the variance of the initial
density field, smoothed on some scale with a specified filter.
We calculate the fraction of the volume which is occupied by peaks which
exceed a threshold value and relate this to the number density of halos of
a specified mass.  The critical density threshold is taken from the theory of
spherical top-hat collapse and is treated as a constant.

\begin{figure*}
\begin{center}
\resizebox{2in}{!}{\includegraphics{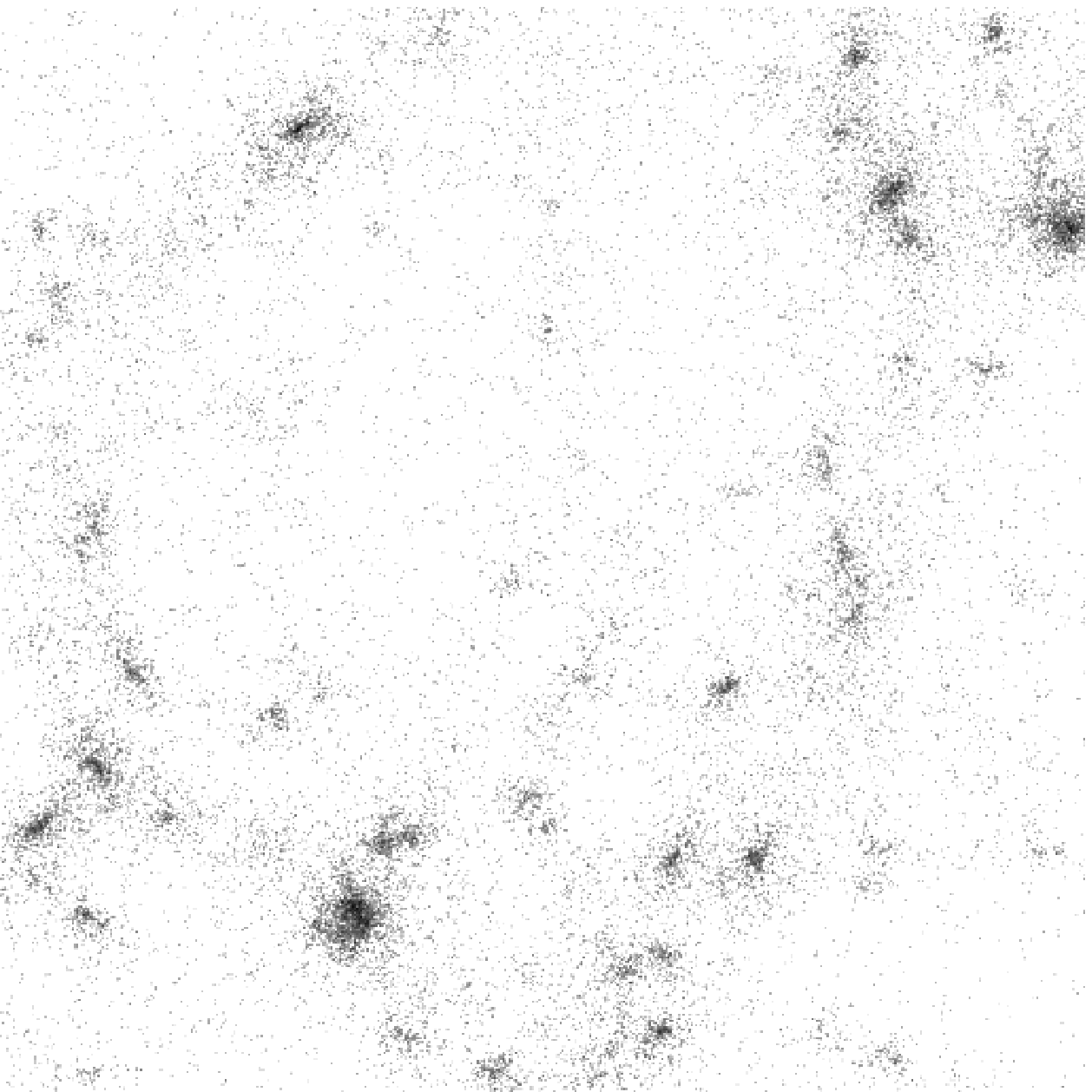}}
\resizebox{2in}{!}{\includegraphics{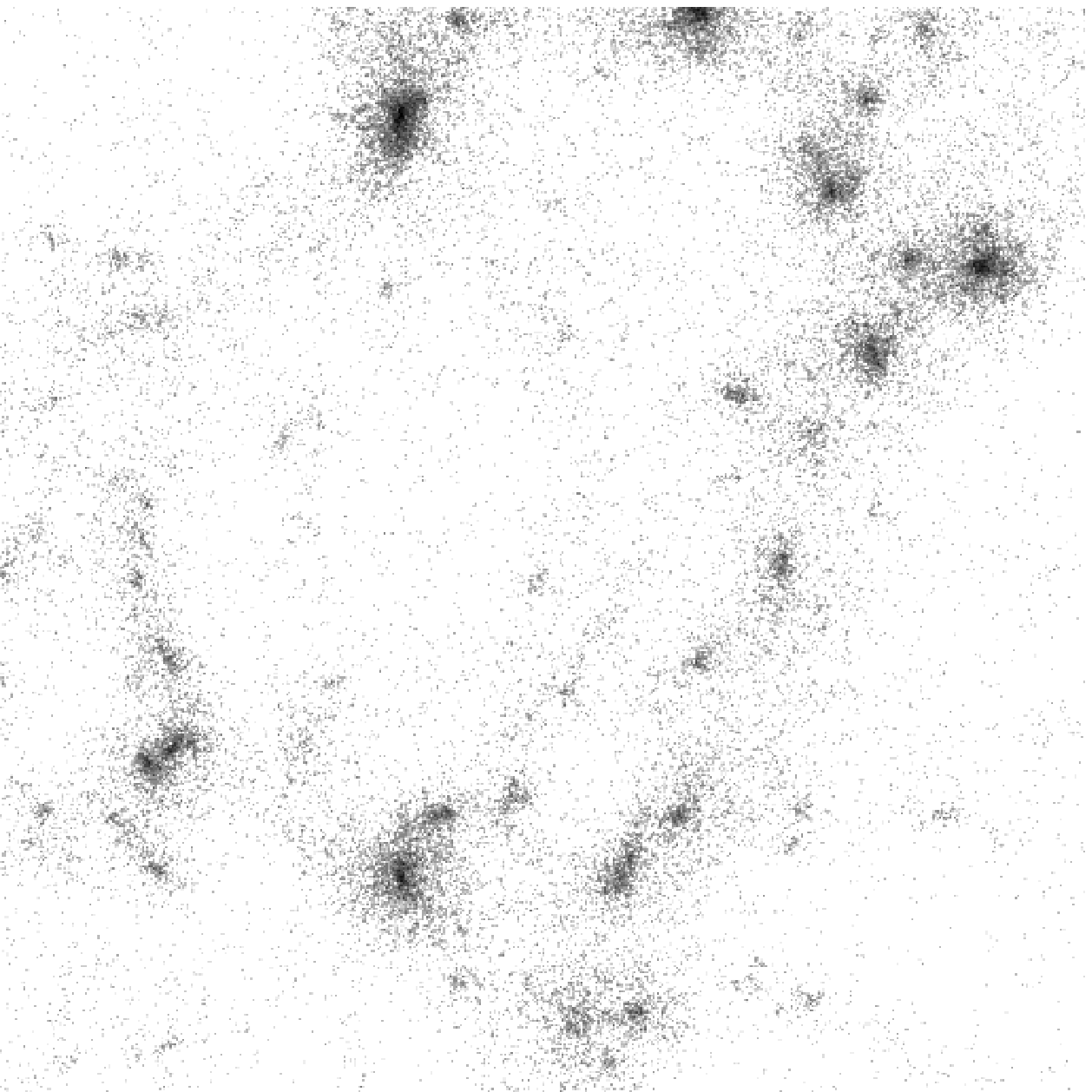}}
\resizebox{2in}{!}{\includegraphics{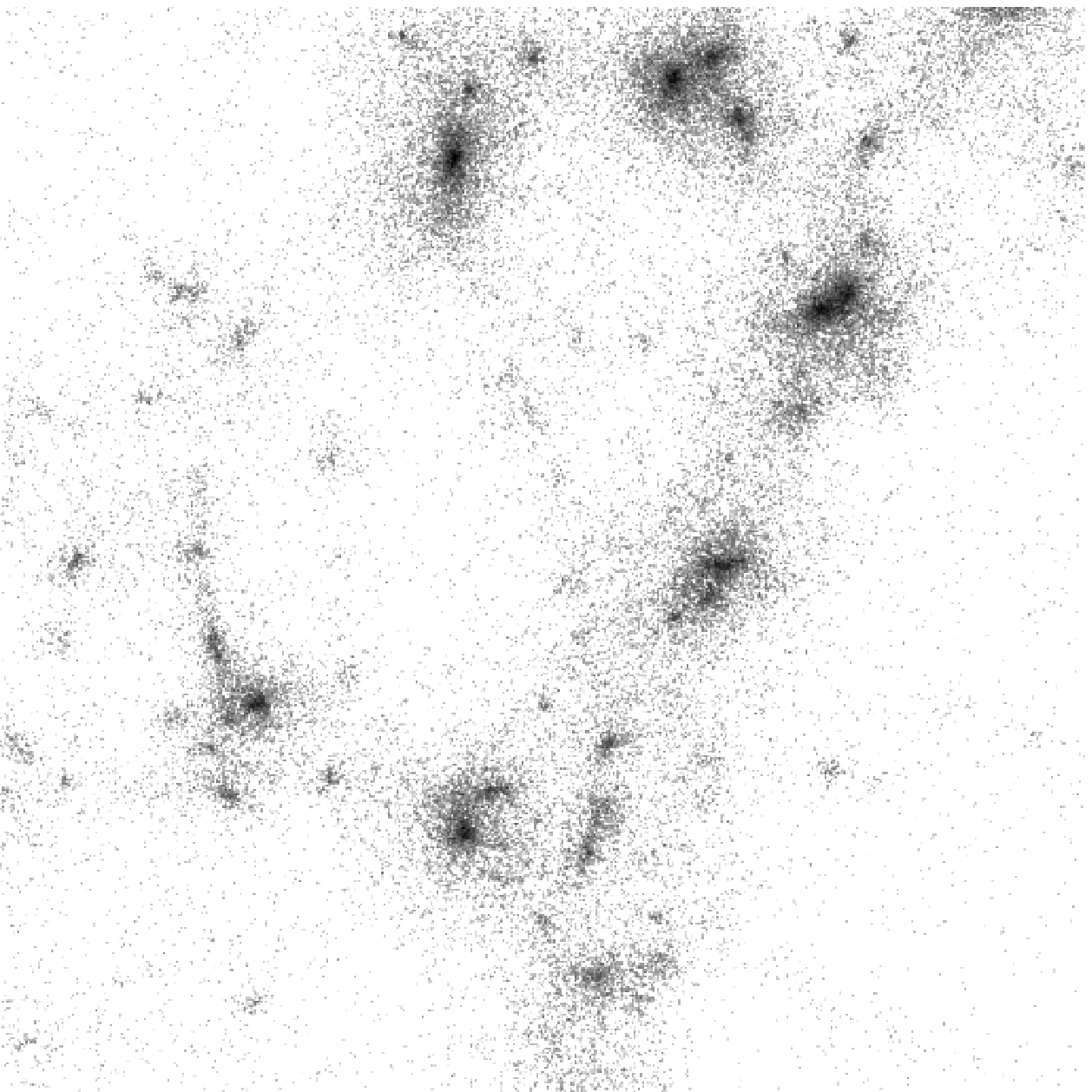}}
\resizebox{2in}{!}{\includegraphics{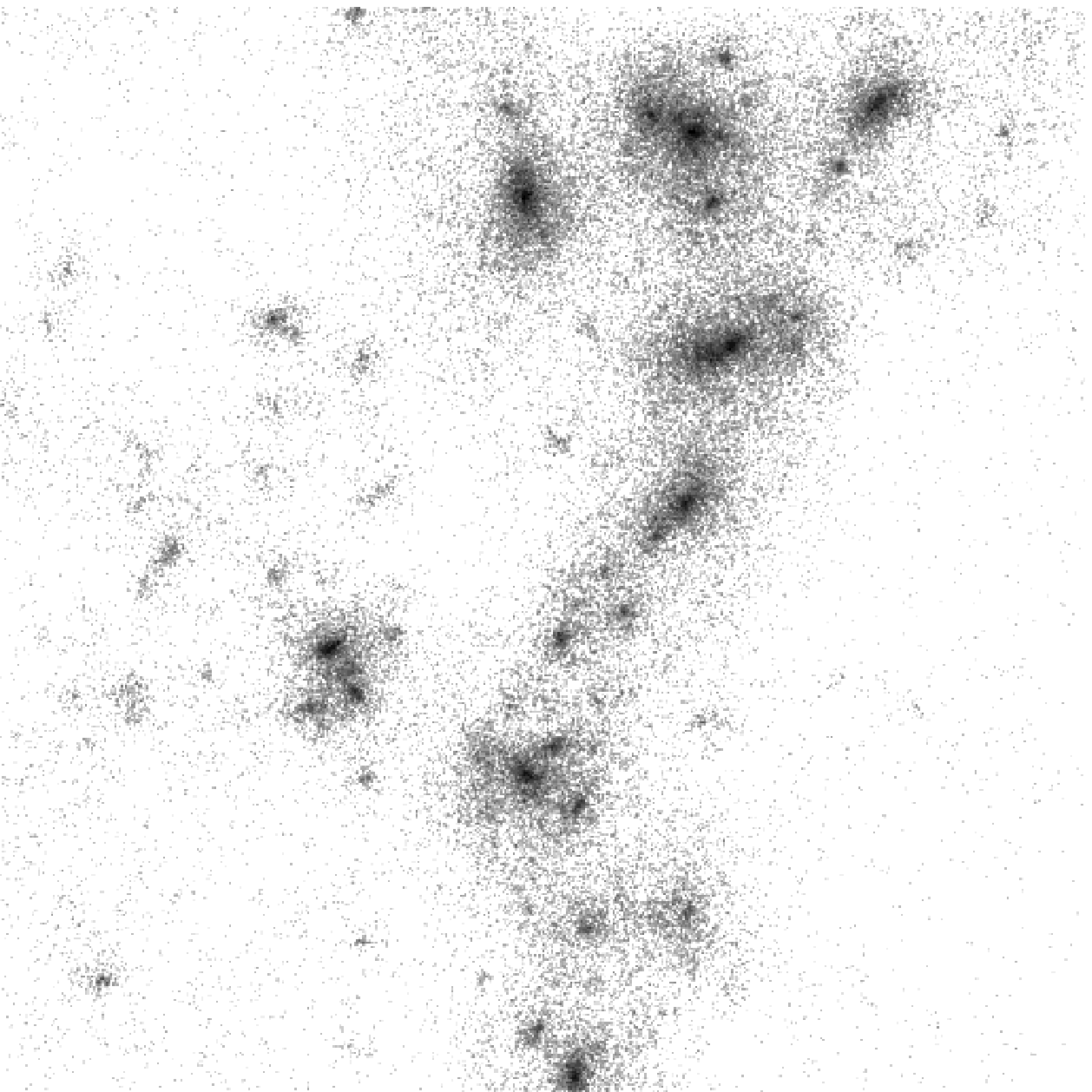}}
\resizebox{2in}{!}{\includegraphics{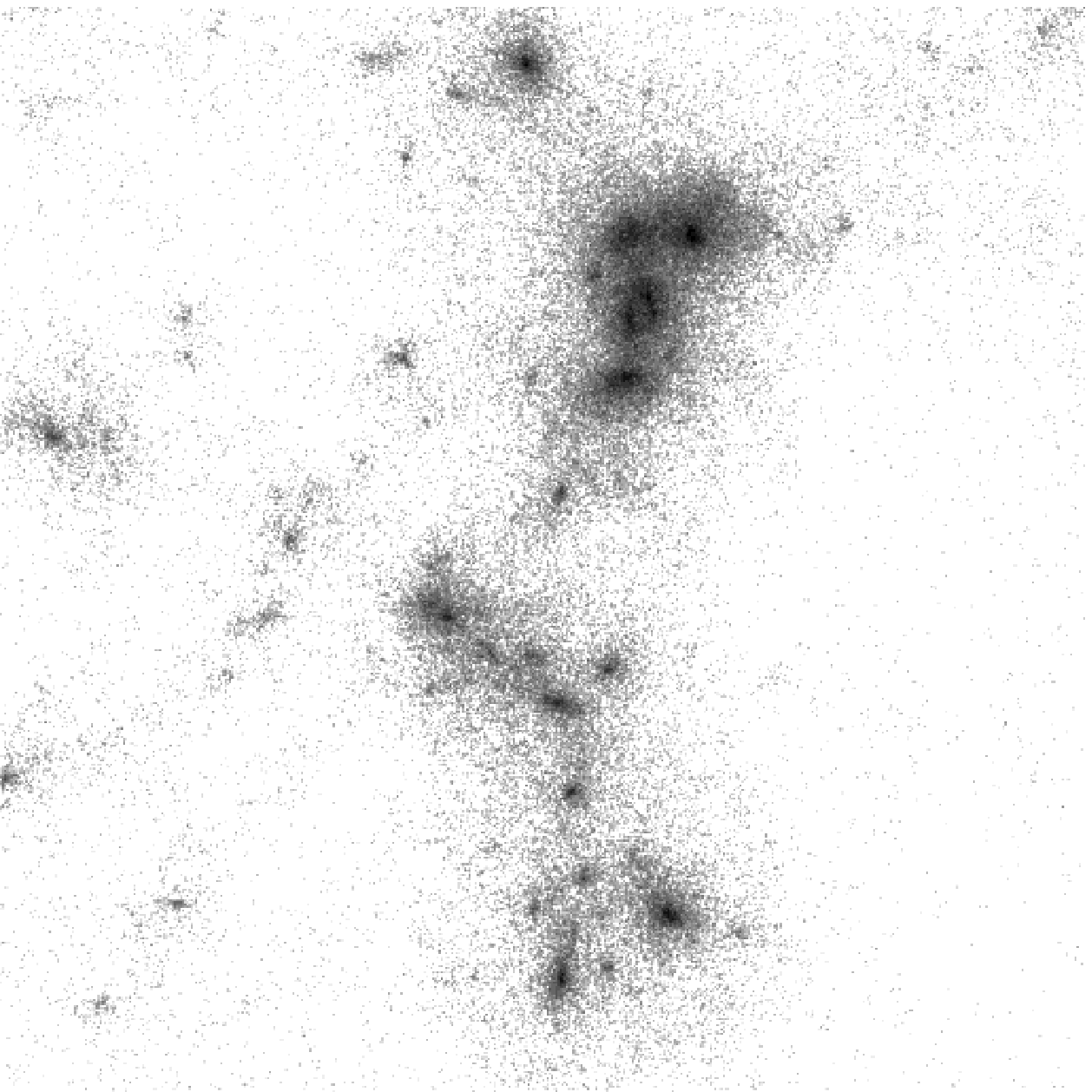}}
\resizebox{2in}{!}{\includegraphics{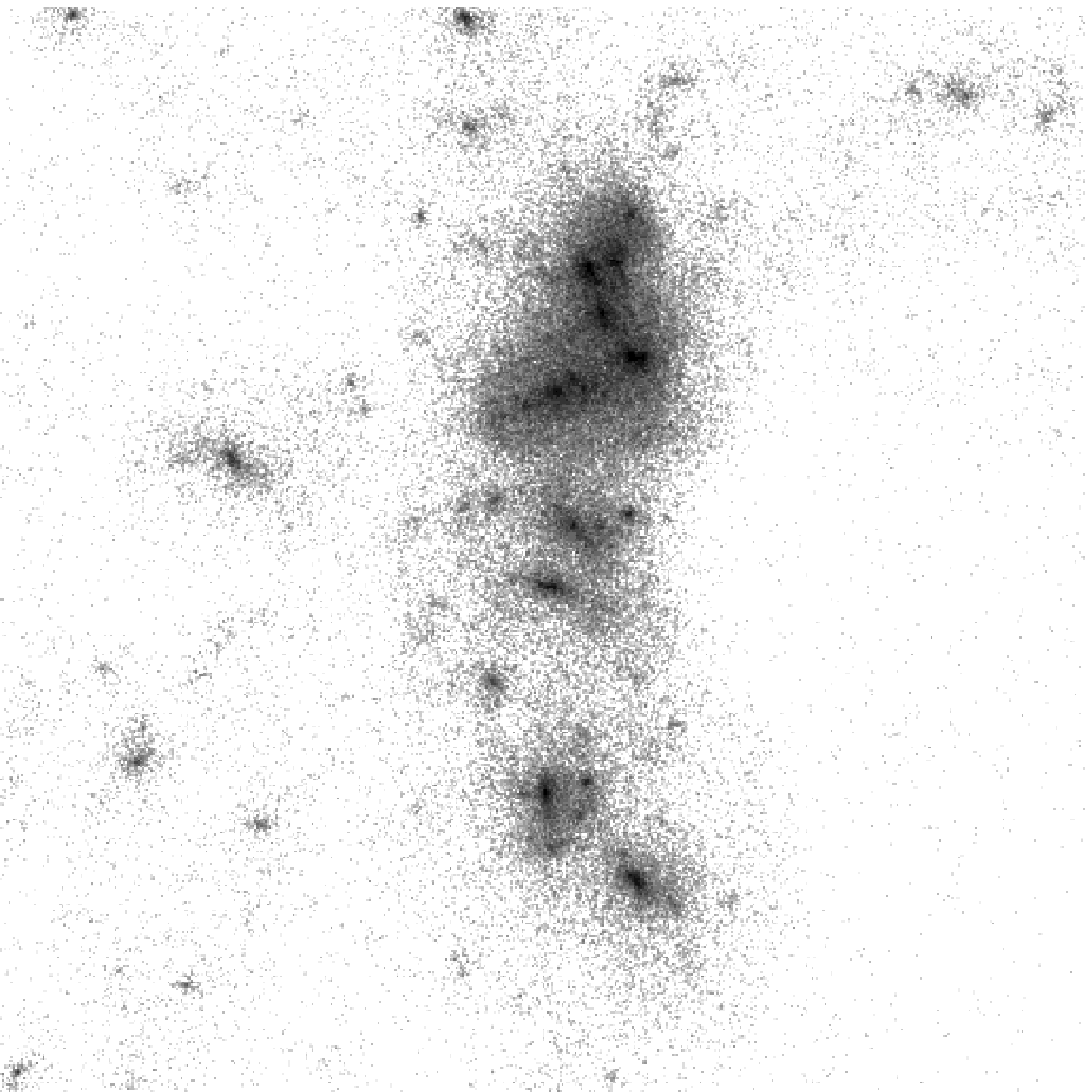}}
\resizebox{2in}{!}{\includegraphics{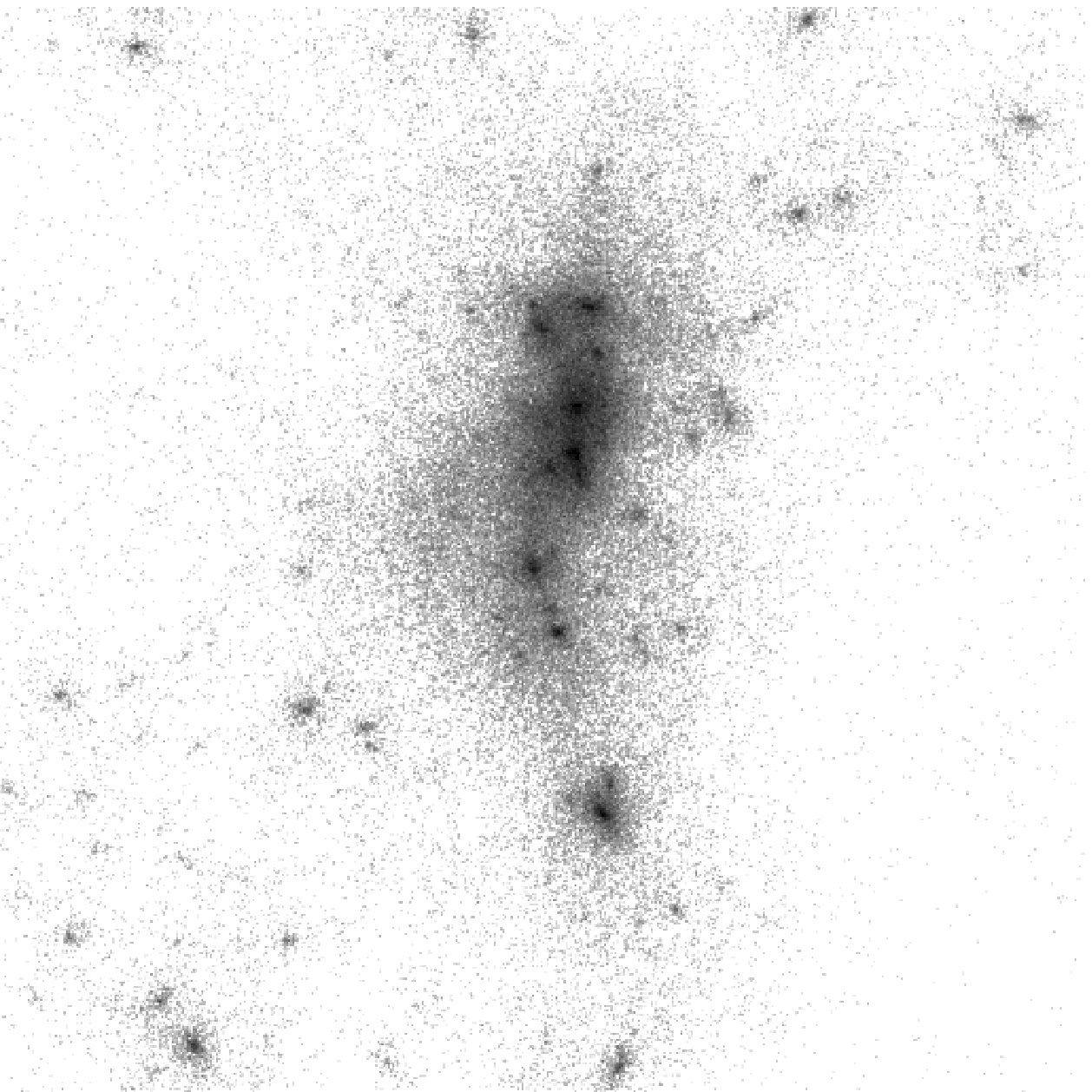}}
\resizebox{2in}{!}{\includegraphics{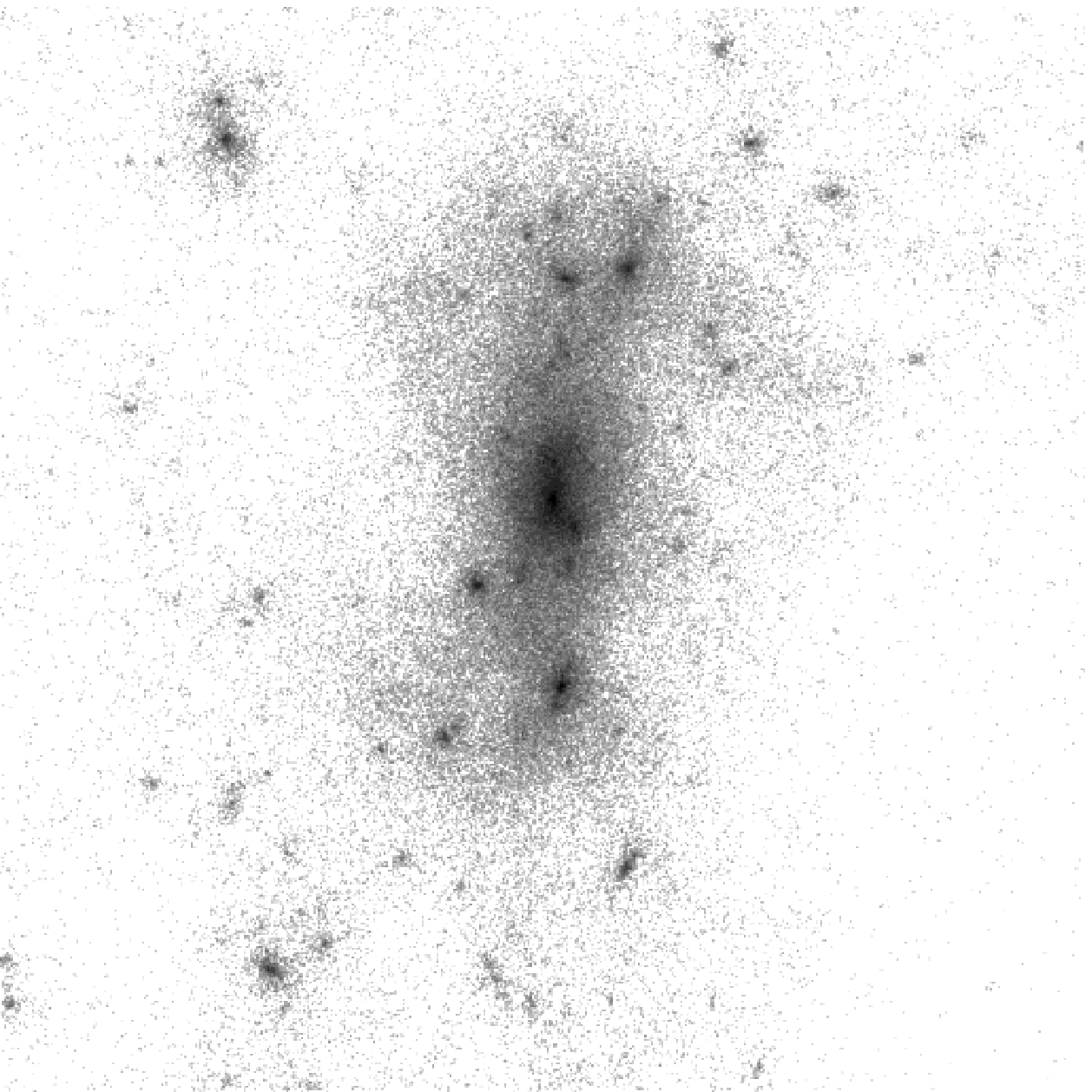}}
\resizebox{2in}{!}{\includegraphics{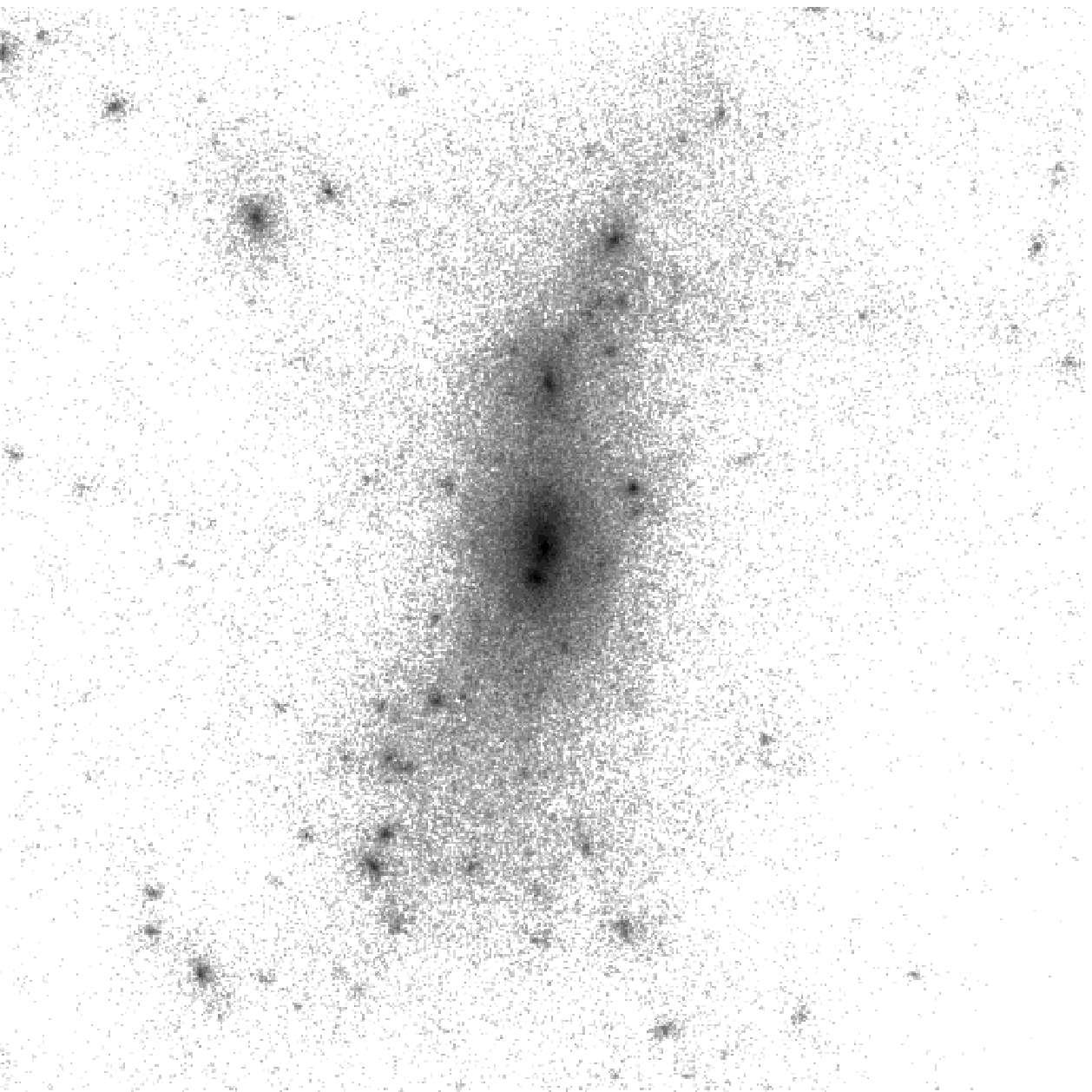}}
\end{center}
\caption{The projected density in a cube $10\,h^{-1}$Mpc on a side centered on
the final position of the second most massive halo in the $512^3$ particle
simulation.  The 9 panels are equally spaced in conformal time from $z=1.5$
to $z=0$.  The grey scale is logarithmic, running from $10^2$ to $10^5$ times
the mean density.}
\label{fig:clusterform}
\end{figure*}

At first sight it seems fortuitous that any result of the complex process of
halo formation within an hierarchical model
(see e.g.~Fig.~\ref{fig:clusterform})
could be derived from the variance of the initial density field, without
reference to any dynamics (see discussion in Lacey \& Cole \cite{LacCol2}).
Or that there should be a `universal' mass function at all.
Indeed JFWCCECY found that the cosmology-independence of the mass function in
scaled units depends upon the mass estimator chosen.  This is a non-trivial
problem because objects in hierarchical models do not have a well defined
outer boundary, making both their identification and the definition of their
total mass convention dependent.
JFWCCECY found best results when using as a mass estimator the sum of the
particle masses in their N-body groups using a particular group finder
(FoF; Davis et al.~\cite{DEFW}, see \S\ref{sec:massdef}).
This differs from the widely followed practice of using the mass within a
spherical region whose radius is derived from the top-hat collapse model
(\S\ref{sec:tophat}).
White~\cite{HaloMass} showed that there is considerable scatter between
the two types of mass estimators which makes it difficult to combine
results which don't use a consistent set.

Is there a middle ground?
The self-similarity of halos observed in simulations has long been taken to
imply that masses are best defined within radii enclosing a fixed density
contrast.  While the density contrast $\Delta_c$ is the conventional choice,
it is not the only possibility.  Since FoF links particles which are
approximately above some density threshold with respect to the background,
the result of JFWCCECY suggests that we should define our masses within fixed
density contrasts with respect to the mean density, not the critical density.
JFWCCECY in fact give such a mass function in their Appendix B.  It uses a
mass within a radius $r_{180b}$ interior to which the mean density is 180
times the background density.  While this is close to the `top-hat' result for
an $\Omega_{\rm m}=1$ cosmology, it extends to extremely large radius
compared to the observable region of clusters.
For example $r_{180b}\simeq 2-3\,h^{-1}\,$Mpc for a rich cluster
($M\sim 10^{15}\,h^{-1}\,M_\odot$).

A number of cosmological tests rely on the existence of a mass function which
is both universal and easy to interpret observationally.  None of the results
described above obviously fulfill these two requirements.
We shall try to make steps towards this goal in the rest of this paper.
Our final solution will be a hybrid which uses the mass estimator $M_{180b}$
suggested by JFWCCECY along with a conversion factor between `observed' and
`theoretical' mass.

\section{Simulations} \label{sec:simulations}

Numerical simulations give qualitative support to the predictions of the
Press-Schechter theory, with small modifications noted at both the high
and low mass ends.  The current state of the art in numerical simulations
aimed at elucidating these departures is the work of JFWCCECY.
Independent confirmations of the JFWCCECY result have been published recently
by White \cite{HaloMass}, Zheng et al.~\cite{ZTWB} and
Hu \& Kravtsov \cite{HuKra}.  In this section we discuss the numerical
simulations we have done to investigate the dependence of the result on
the mass definition chosen.  The reader not interested in the numerical
details is urged to skip to \S\ref{sec:massdef}.

\subsection{N-body runs} \label{sec:nbody}

We have performed a suite of N-body simulations in order to better constrain
the mass/multiplicity function (see Appendix for details of the code).
The first set we used to tune our fitting function, the rest were used as
independent checks.  Throughout we have tried to focus primarily on the
high-mass end of the mass function, which is of the most use for studies of
clusters of galaxies.

Since the primary consideration is one of volume, we have run numerous small
simulations rather than one very large one.  The small simulations were chosen
to have sufficient dynamic range and mass resolution to well resolve a low
mass cluster halo.  Specifically we ran a number of $150^3$ particle
simulations of three different CDM models, each in a $200h^{-1}$Mpc box
(see Table~\ref{tab:sims} for more details).
Each simulation represents a reasonable cosmological volume, so as not to bias
the high mass end of the mass function\footnote{The contribution from modes
with wavelength longer than the box to $\sigma(R)$ in the relevant range is
very small.}, while maintaining enough mass and force resolution to identify
the relevant halos.

\begin{figure*}
\begin{center}
\resizebox{2in}{!}{\includegraphics{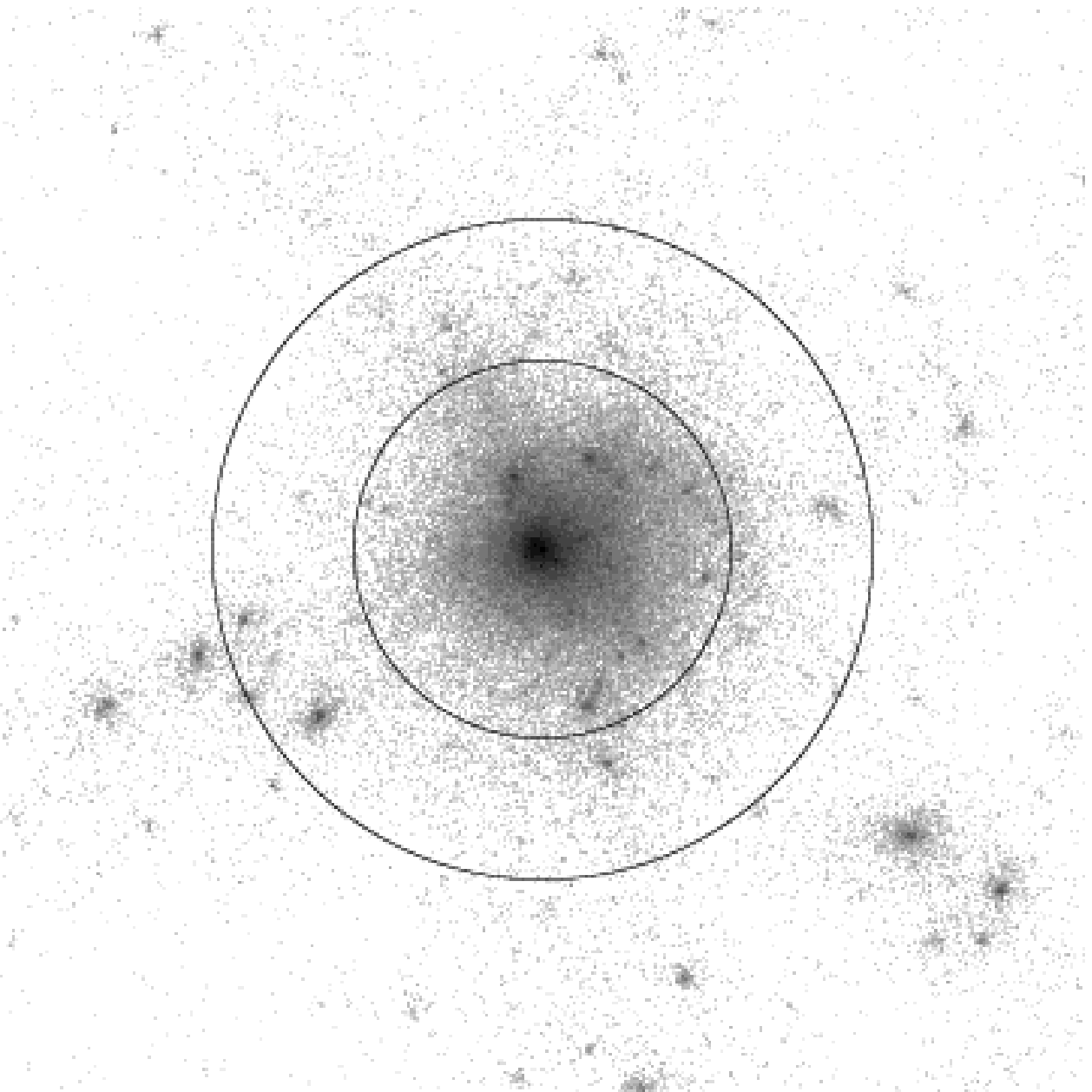}}
\resizebox{2in}{!}{\includegraphics{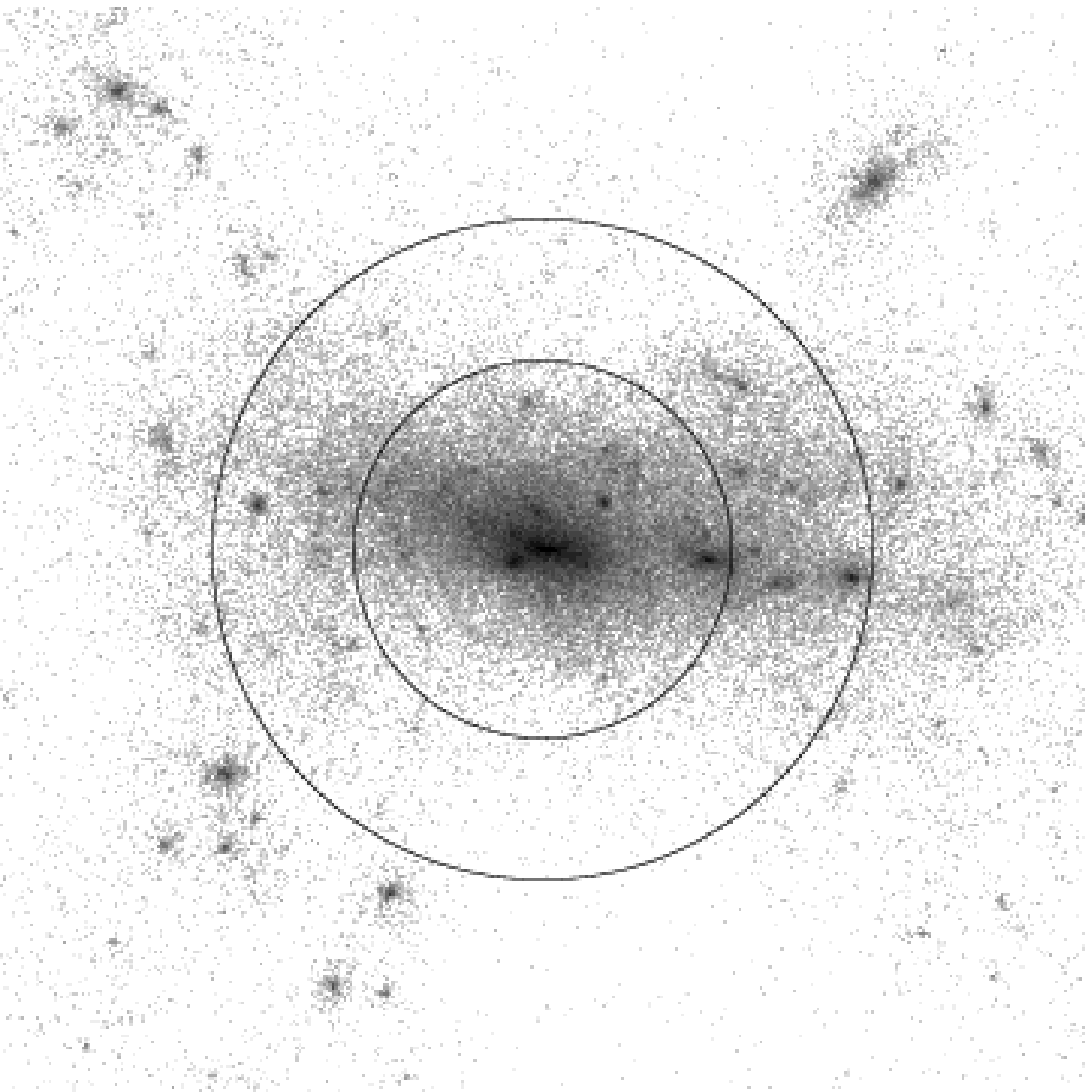}}
\resizebox{2in}{!}{\includegraphics{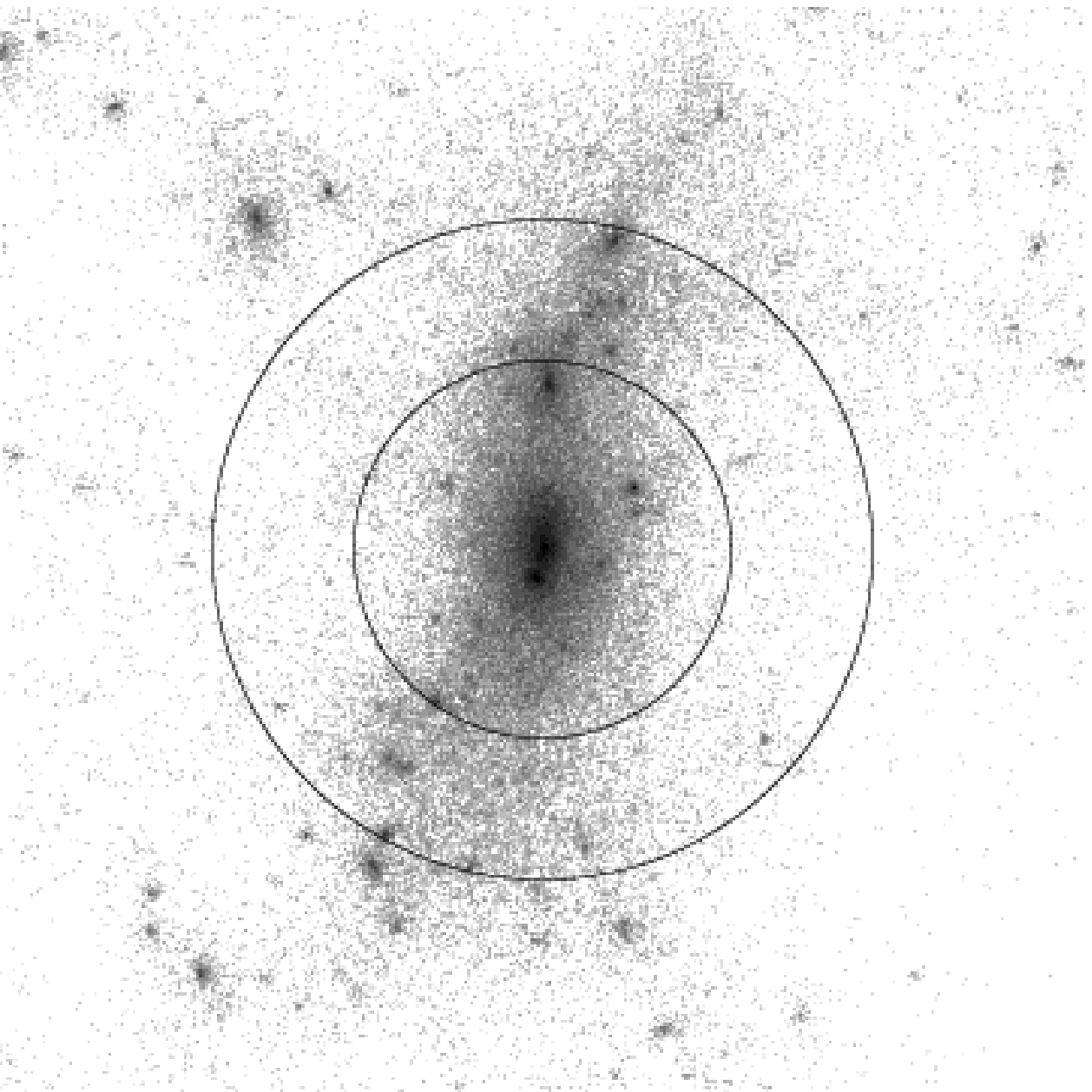}}
\end{center}
\caption{The projected density in a cube $10\,h^{-1}$Mpc on a side centered on
the second most massive halo in the $512^3$ particle simulation.  The 3 panels
are projections down the $x$, $y$ and $z$ axes of the box.
The grey scale is logarithmic, running from $10^2$ to $10^5$ times the mean
density.  The solid circles show $r_{200c}\simeq 1.74\,h^{-1}$Mpc (inner) and
$r_{180b}=r_{54c}\simeq 3.04h^{-1}$Mpc (outer).
Within $r_{180b}$ the material exhibits a wide range of density contrasts.
Note that the halo is neither isolated nor spherical, and has quite a bit of
substructure.}
\label{fig:greyscale}
\end{figure*}

Because it provides a reasonable fit to a wide range of observations,
we first simulated a `concordance' $\Lambda$CDM model which has
$\Omega_{\rm m}=0.3$, $\Omega_\Lambda=0.7$,
$H_0=100\,h\,{\rm km}{\rm s}^{-1}{\rm Mpc}^{-1}$
with $h=0.67$, $\Omega_{\rm B}=0.04$, $n=1$ and $\sigma_8=0.9$
(corresponding to $\delta_H=5.02\times 10^{-5}$).
We call this model 1.
We then changed the model in each of two `orthogonal directions'.
First we changed the mapping between length scale and mass, by changing
$\Omega_{\rm m}$ from 0.3 to 1, while holding the present day power
spectrum fixed (model 2).
Then we changed the normalization of the power spectrum, $\sigma_8$, while
holding the cosmology and shape of the power spectrum fixed (model 3).
Finally we ran a model with a different spectral shape and normalization
as a cross check.  Model 4 had $\Omega_{\rm m}=0.35=1-\Omega_\Lambda$,
$\Omega_{\rm B}h^2=0.02$ and $h=0.65$ with $\sigma_8=0.8$.
Of the models the first has the fewest clusters per unit volume, so
we ran more realizations of this model than the other three.

\begin{table}
\begin{center}
\begin{tabular}{ccccccc}
\hline
Model & $\Omega_{\rm m}$ & $\sigma_8$ &
   $N_{\rm box}$ & $V_{\rm tot}$ & $m_{\rm part}$ & $1+z_{\rm ic}$ \\ \hline
1 & 0.30 & 0.9 & 15 &120 & 1.97 & 40\\
2 & 1.00 & 0.9 & 10 & 80 & 6.58 & 30\\
3 & 0.30 & 1.0 & 10 & 80 & 1.97 & 40\\
4 & 0.35 & 0.8 & 10 & 80 & 2.30 & 40\\
\end{tabular}
\end{center}
\caption{The parameters of the simulations run.  In each case $N_{\rm box}$
different realizations of the Gaussian initial conditions were run.  Each
run used a periodic box of side $200\,h^{-1}$Mpc and $150^3$ particles.
The force softening was of a spline form with a ``Plummer equivalent''
smoothing length of $50\,h^{-1}$kpc -- easily small enough to resolve the
halos of interest.
In the above volumes are quoted in units of $(100\,h^{-1}{\rm Mpc})^3$ and
particles masses in units of $10^{11}\,h^{-1}M_\odot$.}
\label{tab:sims}
\end{table}

We also used an `independent' simulation to check our fitting function.
The cosmology in this case was slightly different than above, having
$\Omega_{\rm m}=0.3$, $h=0.7$ and $\sigma_8=1$.
The simulation employed $512^3$ particles in a $300\,h^{-1}$Mpc with a
smoothing length of $20\,h^{-1}$kpc.  We can use this simulation to see
whether the mass function extrapolates correctly to lower masses
(where it becomes a power-law), see \S\ref{sec:universal}.

\subsection{The group catalogs}

{}From the $z=0$ output of each simulation we produce a halo catalogue by
running a ``friends-of-friends'' group finder (e.g.~Davis et al.~\cite{DEFW})
with a linking length of either $b=0.2$ (in units of the mean interparticle
spacing) or $b=0.1$.  We can use these two different group catalogs to test
the sensitivity of our results to the selection of halos.
The FoF algorithm partitions the particles into equivalence classes, by linking
together all particle pairs separated by less than a distance $b$.
We keep all groups above 32 particles, which imposes a minimum halo mass of
order $10^{13}h^{-1}M_\odot$.
[FoF groups with more than 32 particles are known to be robust.]
The FoF algorithm as we have defined it cannot be used to address sub-structure
in the halos that we find, but for our purposes this will not be a serious
limitation as the P-S formalism also completely neglects sub-structure.

Several other group finders exist which we could have used in addition to
the FoF algorithm.  Some of these begin with groups defined in a FoF manner
while some find a partition of the particles in a completely different way.
Luckily, an exploration of all of these different group finders will turn out
to be unneccesary.  For most of the mass estimates defined in
\S\ref{sec:massdef} the precise group finding algorithm is unimportant.
We shall show later that the mass functions obtained with two different FoF
groups, with radically different partitionings of the particle distribution,
are almost identical.
This may be telling us that the physical properties of the group which we are
calculating, the `total mass', are independent of the details of how the group
is originally found as an overdensity in three dimensional space.

In order to define the mass it will be very useful to have a halo center.
We define the center of a halo as the position of the potential minimum,
calculating the potential using only the particles in the FoF group.  This
proved to be more robust than using the center of mass, as the potential
minimum coincided closely with the density maximum for all but the most
disturbed clusters.  Additionally, the position of the center was very
insensitive to the presence or absence of the particles near the outskirts
of the halo, and thus to the precise linking length used.

\section{The mass of a halo} \label{sec:massdef}

As remarked earlier, since the objects formed in a hierarchical model have
no clear boundary, all mass definitions are a matter of convention.
For each halo in each catalog we computed 8 different definitions of mass.
As we shall see later, it is only those definitions which encompassed the
majority of the virialized material (\S\ref{sec:virial}) which will turn
out to be useful for our purposes, and the best shall be $r_{180b}$.

First we used the `FoF mass', simply the number of particles in the group
times the particle mass.  With $b=0.2$, this definition is the one used by
JFWCCECY, for which they found a universal mass function.
By considering the mean number of particles in a sphere of radius $b$ one
can argue that (if all particles have the same mass) FoF groups are bounded
by a surface of density $3/(2\pi b^3)\simeq 60$ times the background.
If all groups were spherical and singular isothermal spheres\footnote{If
$\rho(r)\propto r^{-2}$ the mean density interior to a radius where the
density is $\rho$ is just $3\rho$.}, this would imply a mean density inside
the FoF group of roughly $180$ times the background density or
$180\Omega_{\rm m}$ times the critical density.  In practice there is a very
large scatter about this value.
We also use the sum of the particles in the $b=0.1$ groups.

Motivated by the self-similarity exhibited by halos in simulations we also
define the mass from a spherically averaged profile about the cluster center.
Specifically we define $M_\Delta$ as the mass contained within a radius
$r_\Delta$ inside of which the mean interior density is $\Delta$ times the
{\it critical\/} density
\begin{equation}
  \int_0^{r_\Delta} r^2 dr\ \rho(r) =
    {\Delta\over 3} \rho_{\rm crit} r_\Delta^3
  \qquad .
\end{equation}
The `virial mass' from the spherical top-hat collapse model would then be
simply $M_{\Delta_{\rm c}}$.  We shall refer to this mass as
$M_{{\rm th-vir}}$ rather than the somewhat vague term virial mass.
Since $\Delta_{\rm c}\simeq 200$ in a critical matter density cosmology,
many authors mean by `virial mass', $M_{200}$.  This is in fact the most
common definition.  We shall write this $M_{200c}$ to make explicit the
fact that it is with respect to the critical density.
We shall also consider $M_{500c}$, which has the advantage that it probes
material at sufficiently small radii that it is often directly accessible to
X-ray observations.

In the above we have followed common usage and measured the mean interior
density contrast to the critical density.  This has historically been motivated
by (a) considerations based on the virial theorem, which provides estimates of
the halo velocity dispersion and `temperature', in which the critical density
provides a natural scale and (b) because it requires no assumptions about the
cosmological parameters (beyond $h$) in its definition.
More recent work, specifically JFWCCECY, has suggested that the halo mass
function may be universal if masses are measured within a fixed density
contrast measured not with respect to the critical density, but with respect
to the {\it background\/} density.
We shall follow their lead and also calculate $r_{180b}$, where the mean
interior density is $180$ times the background density.
Note that while the mass function may be more universal with this definition,
it comes with an associated price from an observational point of view:
it introduces a dependence on an assumed $\Omega_{\rm m}$ in the definition.

Unfortunately both the FoF mass with $b=0.2$ and $r_{180b}$ encompass a very
wide range of material (see Fig~\ref{fig:greyscale}), far beyond what can
usually be observed and into the region where the profiles start to show
significant scatter from halo to halo.
For a cluster mass halo, $r_{180b}$ can be 75\% larger than the most widely
used $r_{200c}$.
For this reason we shall also consider $M_{500b}$ and $M_{1000b}$, which
require less of an extrapolation.

In cases where we use a small linking length and a large $r_\Delta$ it is
possible that two halos overlap.  Since our definition of mass for each halo
includes all of the mass within $r_\Delta$, not just that associated with the
FoF halo itself, this can result in us double-counting the mass in the overlap
region.  This is an unfortunate side-effect of a mass estimator based on
spherical averages for objects which are neither spherical nor isolated.
To avoid this we cull from each run the smaller of two halos whose centers are
closer than the sum of the `virial' radii.  This procedure is then relatively
insensitive to the linking length used to define the original halos.  If we
had used a larger linking length and `merged' the extra halo with the larger
one (removing it from out list), the mass assigned to the combined halo would
still be that within $r_\Delta$ of the potential minimum of the combined group
-- presumably the potential minimum of the larger mass group -- and thus be
the mass originally assigned to the larger of the two groups.
If we do not perform this culling step then we find a significant excess of
low mass objects compared to the analytic predictions.  With the culling the
mass function becomes almost totally insensitive to the original group finder
parameters.

The need for this step is more than just a technical issue.  It stems from the
fact that in hierarchical models halos are rarely isolated, but are often
found in various stages of merging or accretion.  Observers often remove from
their samples systems which they deem to be interacting too strongly or too
recently.  This can introduce a `bias' in the mass function.
Rather than attempt to quantify how `disturbed' various halos are or whether
there is observational evidence for interaction which would cause them to be
removed from any particular sample, we have chosen to apply a criterion that
can be equally well applied in simulations and observations:
we omit the smaller of any two systems whose virial radii overlap.
The fraction of the halos culled in each of the models is relatively small.
For example in Model 1 for the $b=0.2$ halos the culled fraction is less than
1\% for all of the mass definitions.  For $b=0.1$ it is 15\% for $M_{180b}$,
$8\%$ for $M_{\rm th-vir}$ and around 1\% or less for the other mass
definitions.
How closely our treatment mimics the selection of individual objects in
current samples is unclear.  As we demand increasing accuracy in our
comparison between theory and observation this issue will need to be revisited.

\section{The virial region} \label{sec:virial}

Our definition of a halo is primarily one of density contrast.  An alternative
definition is that a halo contains material which has broken away from the
universal expansion and is (at least approximately) in virial equilibrium.
As we shall see below, it is only for density thresholds where these two
definitions roughly coincide that one obtains a nearly universal mass function.
Since the virial region encompasses such a large volume of space, this will
require us to investigate extrapolations from the inner, easier to measure,
regions.

\begin{figure}
\begin{center}
\resizebox{3.5in}{!}{\includegraphics{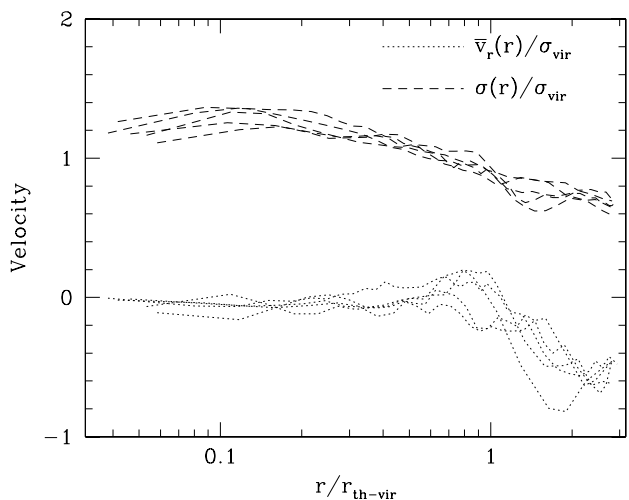}}
\end{center}
\caption{The radial velocity (dotted) and velocity dispersion (dashed)
profiles of the 5 most massive clusters in the $512^3$ simulation described
in the text.  Each cluster has ${\cal O}(10^5)$ particles and
$M_{200c}>10^{15}\,h^{-1}\,M_\odot$.  Points are plotted every 5000 particles
in radius out to $3\times$ the top-hat virial radius.  The velocities are
all normalized to the 3D velocity dispersion of the dark matter within
$r_{\rm th-vir}$.}
\label{fig:virial}
\end{figure}

As a first step it is interesting to ask how well the spherical top-hat
collapse model approximates the messy formation process of a cluster in a
hierarchical model (see Fig.~\ref{fig:clusterform}).
To investigate this we have looked at the five most massive clusters in
the $512^3$ particle simulation described earlier.
For each cluster we calculate the mean radial velocity, $\bar{v}_r(r)$,
and the velocity dispersion, $\sigma(r)$, in bins containing 5000
particles from the center out to 3 times the virial radius predicted by
the top-hat model ($r_{101}$ for $\Omega_{\rm m}=0.3$).
The results are shown in Fig.~\ref{fig:virial}.

Fig.~\ref{fig:virial} shows that the clusters are roughly isothermal, with a
velocity dispersion profile that peaks near the break radius (where the
density profile has a slope of $-2$).
Inside of the virial radius the mean velocity is close to zero, in units
of the 3D velocity dispersion $\sigma_{\rm vir}$.
Just outside the virial radius the transition from `virialized' material
to in-flowing material ($\bar{v}_r<0$) is clearly seen in all 5 clusters.
Thus it seems that the `virial radius' is predicted within a factor of 2 by
the top-hat model, though the clusters exhibit some scatter.
The radius $r_{180b}$ is only 30\% larger than $r_{\rm th-vir}$ for a
cluster mass halo in this cosmology, so within the scatter we could take
the `virial' radius to be $r_{180b}$ also.
Using a higher density contrast than $\Delta_c$,
e.g.~$M_{200c}$ or $M_{500c}$,
would clearly underestimate the transition radius for all of the clusters.

\begin{figure}
\begin{center}
\resizebox{3.5in}{!}{\includegraphics{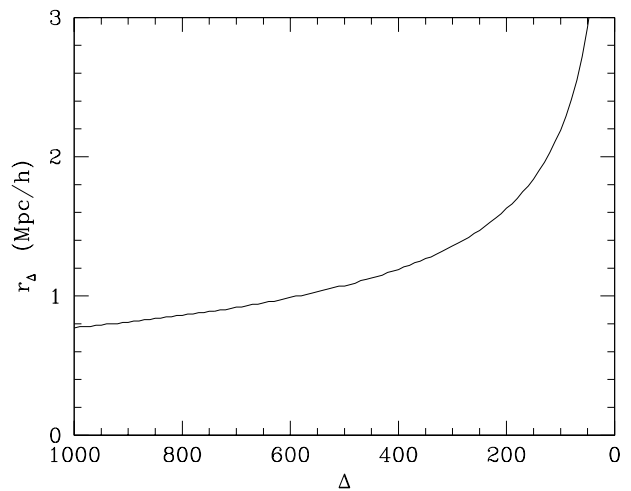}}
\end{center}
\caption{The radius $r_\Delta$ within which the mean density is $\Delta$
times the {\it critical\/} density for an NFW halo with
$M_{200c}=10^{15}\,h^{-1}M_\odot$ and $c=5$.
For $\Omega_{\rm m}=0.3$ the `universal' density contrast of $r_{180b}$
is at $r_{54c}\simeq 2.8\,h^{-1}$Mpc.}
\label{fig:rdelta}
\end{figure}

To get a feel for the translation between density contrast and `size' for
a rich cluster, we can make use of the universal density profile of
Navarro, Frenk \& White~\cite{NFW}.  These authors defined the virial radius
as $r_{200c}$ and density contrasts with respect to critical.  The universal
form of the density profile is
\begin{equation}
  {\rho(r)\over\rho_{\rm crit}} \propto {1\over x(1+x)^2}
\end{equation}
where $x=r/r_s$ is a scaled radius and $r_s$ describes the transition from
$r^{-1}$ to $r^{-3}$ in the profile.  We show the radius within which the
mean density is $\Delta$ times the critical density in Fig.~\ref{fig:rdelta}
for a rich cluster with $M_{200c}=10^{15}\,h^{-1}M_\odot$ and a concentration
$c\equiv r_{200c}/r_s=5$.
Note that for $\Omega_{\rm m}=0.3$, $r_{180b}=r_{54c}\simeq 2.8\,h^{-1}$Mpc.
For lower $\Omega_{\rm m}$ it would exceed $3\,h^{-1}$Mpc.

\section{A universal mass function?} \label{sec:universal}

\begin{figure*}
\begin{center}
\resizebox{6in}{!}{\includegraphics{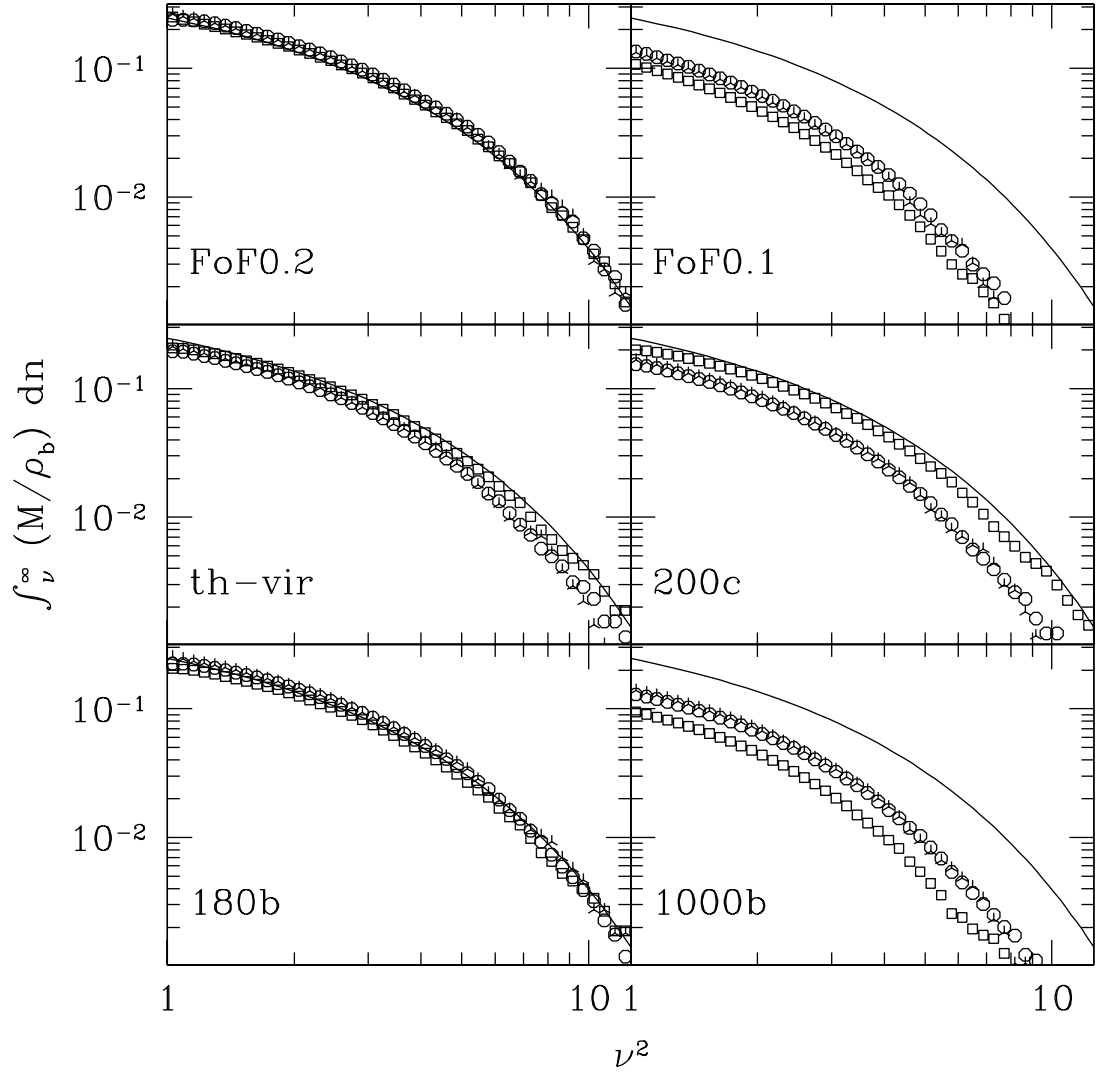}}
\end{center}
\caption{The multiplicity function vs.~the peak height $\nu^2$ for our 3
models and for 6 of our 8 mass definitions.
Open circles are Model 1, open squares Model 2 and triangles Model 3.
The solid line is the S-T fitting function.
The panels are labeled with the halo mass definition used.
This indicates to what extent the mass functions are indeed universal.}
\label{fig:universal}
\end{figure*}

Given a (possibly culled) set of halos each with a known mass, we wish to
find a fitting function to the mass/multiplicity function.
As a first step we check whether the mass functions do indeed form a
`universal' multiplicity function by plotting
\begin{equation}
 N(>\nu) \equiv \int_\nu^\infty  {M\over\bar{\rho}} {dn\over d\nu} d\nu
\end{equation}
vs.~the peak height $\nu^2$ (Fig.~\ref{fig:universal}).
As we can see, for the 3 models shown\footnote{We omit here the results of
Model 4 for visual clarity.  This Model will be reinstated in some later
plots.}, it is a good approximation to assume that all of these mass functions
come from a universal multiplicity function for the FoF halos with $b=0.2$.
The universal form is quite well fit by the Sheth \& Tormen form of
Eq.~(\ref{eqn:fnu}).  This confirms the earlier work of JFWCCECY.

The mass function is also close to universal if the top-hat virial mass is
used, though the agreement is not as good as in the $b=0.2$ case and would
need to be checked for a wider range of cosmologies.
As noted by JFWCCECY the mass $M_{180b}$ also gives a close to universal mass
function.
Each of these mass estimators includes the majority of the mass within the
virialized region (see Fig.~\ref{fig:virial}).

As expected, the mass function within $r_{200c}$ shows a systematic difference
between the $\Omega_{\rm m}=1$ model and the other two.  This is because the
mass is defined interior to a density contrast which doesn't scale with the
background density.

Of particular interest is the case of $M_{1000b}$.  Here the mass is defined
in terms of a density contrast with respect to the background density (like
the `universal' $M_{180b}$), but at a higher density.  This region of the
halo is more amenable to observation.
However we note that as we increase the density threshold, focusing on the
inner regions of the halos, the universality of the mass function degrades.
[The case $M_{500b}$ is intermediate, and is omitted from the figure.]

\begin{figure}
\begin{center}
\resizebox{3.5in}{!}{\includegraphics{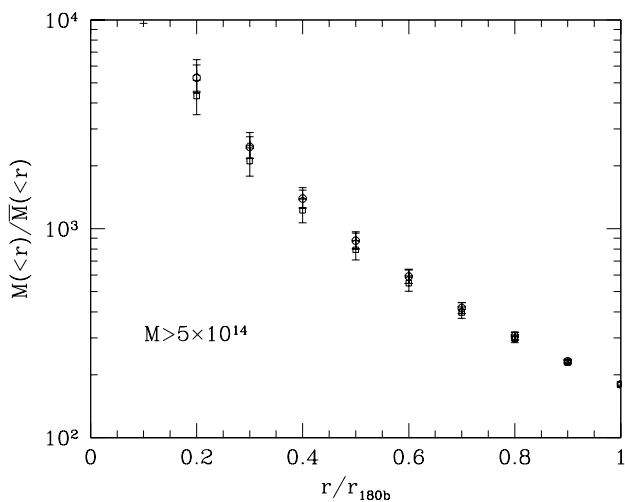}}
\end{center}
\caption{The (interior) mass profiles in scaled units for the halos above
$M_{180b}=5\times 10^{14}\,h^{-1}\,M_\odot$ in the 3 models.  As before,
open circles are Model 1, squares Model 2 and triangles Model 3.
The mean and standard deviation of the profiles is shown.  Note that by
definition $M(<r_{180b})=180\bar{M}(<r_{180b})$ so there is no scatter in
this point.}
\label{fig:mprof}
\end{figure}

\begin{figure}
\begin{center}
\resizebox{3.5in}{!}{\includegraphics{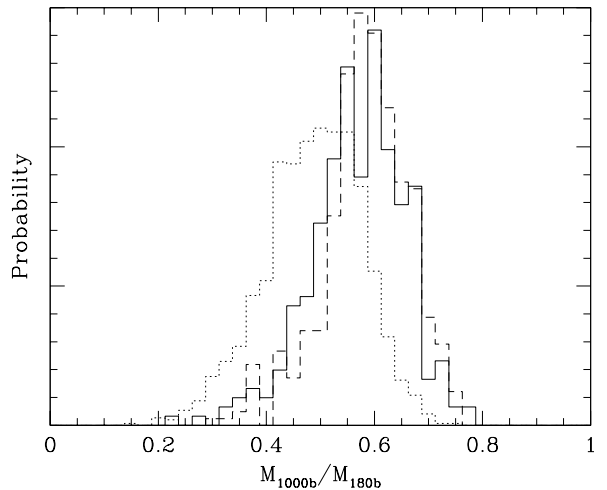}}
\end{center}
\caption{The ratio of different mass estimators for clusters with
$M_{180b}>5\times 10^{14}\,h^{-1}\,M_\odot$ in the 3 models.
Solid line is Model 1, dotted Model 2 and dashed Model 3.}
\label{fig:mratio}
\end{figure}

We can understand this result by considering the different formation
histories of the halos.  Fig.~\ref{fig:mprof} shows the average mass profiles
of the most massive halos from the 3 models in scaled units.
The halos in Model 2, with $\Omega_{\rm m}=1$, form later and thus are less
concentrated than the halos in Models 1 and 3.  This increases the ratio
$M_{1000b}/M_{180b}$ (see Fig.~\ref{fig:mratio}) or $M_{1000b}/M_{200c}$.
On an object by object basis this scatter is fairly large.
This is unfortunate since it is precisely the inner regions which are most
amenable to observation!
Thus the mass estimators for which the mass function is close to universal are
those which require an extrapolation beyond the observed region
(out to $2-3\,h^{-1}$Mpc), and this extrapolation is quite sensitive to
both the cosmology and the particular formation history of the halo under
consideration.

\begin{figure}
\begin{center}
\resizebox{3.5in}{!}{\includegraphics{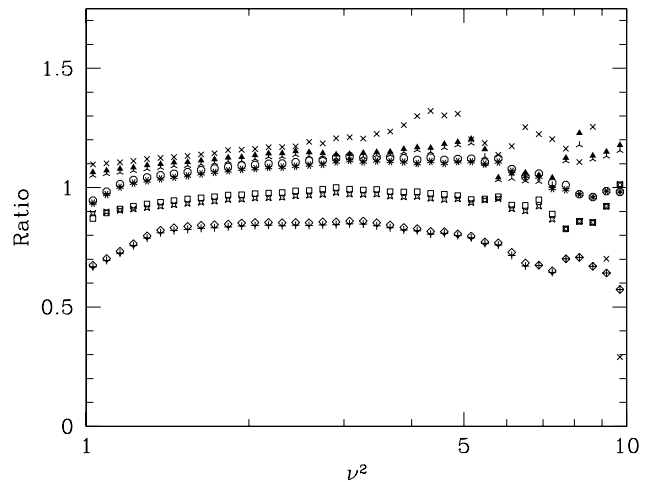}}
\end{center}
\caption{The multiplicity function vs.~the peak height $\nu^2$, divided
by the fitting function of Eq.~(\protect\ref{eqn:fnu}), for $M_{180b}$.
Open circles and stars are Model 1 with $b=0.2$ and $b=0.1$ respectively.
Open squares and 3-pointed crosses Model 2, triangles Model 3 and
plusses and diamonds Model 4 (with worse statistics at the high mass end).
The 4-pointed crosses are the results from the $512^3$ run of a slightly
different model (see text).
At the high mass end ($\nu\to\infty$) our statistics become very poor.}
\label{fig:unilin}
\end{figure}

\begin{figure}
\begin{center}
\resizebox{3.5in}{!}{\includegraphics{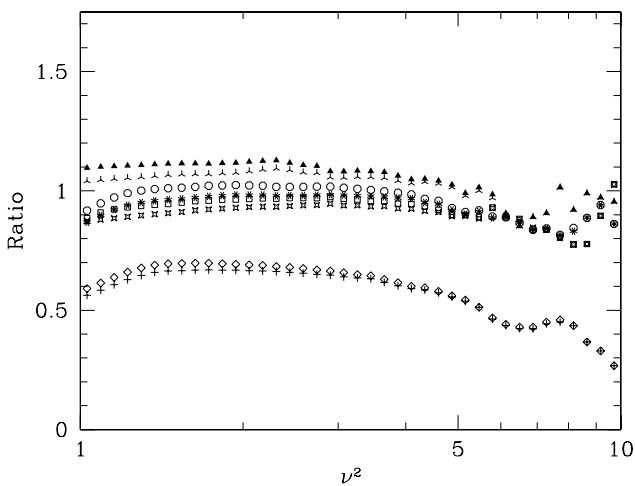}}
\end{center}
\caption{The multiplicity function vs.~the peak height $\nu^2$, divided
by the fitting function of Eq.~(\protect\ref{eqn:fnu}) as above.
Here we have converted from $M_{500c}$ to $M_{180b}$ assuming the halos
are all of the NFW form with $c=5$.
Open circles and stars are Model 1 with $b=0.2$ and $b=0.1$ respectively.
Open squares and crosses Model 2, triangles Model 3 and plusses and
diamonds Model 4.}
\label{fig:mconv}
\end{figure}

We show in Fig.~\ref{fig:unilin} the mass functions on a linear scale.
We report the results as a ratio of the N-body results to the integral of
Eq.~(\ref{eqn:fnu}) for the case of $M_{180b}$.
Each of the four models is represented by two sets of points, one where the
halos are initially found with a linking length of $b=0.2$, and the other
with $b=0.1$.  As we can see there is almost no dependence on the initial
group finder used in this case.
As a cross check we also show on this linear scale the mass function from
the $512^3$ particle simulation in the $300\,h^{-1}$Mpc box.  This run has
much higher mass and force resolution than the majority of the runs used
here, and a slightly different cosmology i.e.~spectral shape and normalization
on the scales of interest.  The total volume is however smaller, so the
statistics at the $\nu\to\infty$ end are much poorer.
However it makes a good `independent' check of the universality of the mass
function.
The scatter between the models is at the $\pm 20\%$ level.

We also show, on the same linear scale, the results of converting between
different spherically averaged mass profiles in Fig.~\ref{fig:mconv}.
Following White \cite{HaloMass} we show in particular what happens if one
constructs the mass function by measuring $M_{500c}$, converting this to
$M_{180b}$ using an NFW profile with $c=5$.
We make no correction for the large scatter we saw in the detailed comparisons
above, we simply apply a numerical rescaling.
For $\Omega_{\rm m}=0.3$ the conversion is $M_{180b}\simeq 2.0M_{500c}$ while
for $\Omega_{\rm m}=1$ the conversion is $M_{180b}\simeq 1.4M_{500c}$.
As we can see, while the conversion shows a lot of scatter it introduces no
major bias and the mass functions so constructed are close to universal.
Such a procedure was followed, in reverse, by Pierpaoli et al.~\cite{PSW} and
Seljak \cite{Sel} for example.
A more complicated conversion along the same lines has been provided by
Hu \& Kravtsov \cite{HuKra}.
This is very encouraging because it means that, while $M_{180b}$ cannot
reliably be measured on an object-by-object basis, a noisy estimator of it
can be easily constructed which turns out to be good enough to construct the
`universal' mass function.  The outlier is Model 4 which was already known to
be discrepant in Fig.~\ref{fig:unilin}.  The rescaling has made it more
discrepant from the mean, indicating that this procedure is not without its
flaws, but even so the mass function is predicted to 30\% over much of the
range.

\section{Fitting the mass function} \label{sec:fitting}

\begin{figure*}
\begin{center}
\resizebox{6in}{!}{\includegraphics{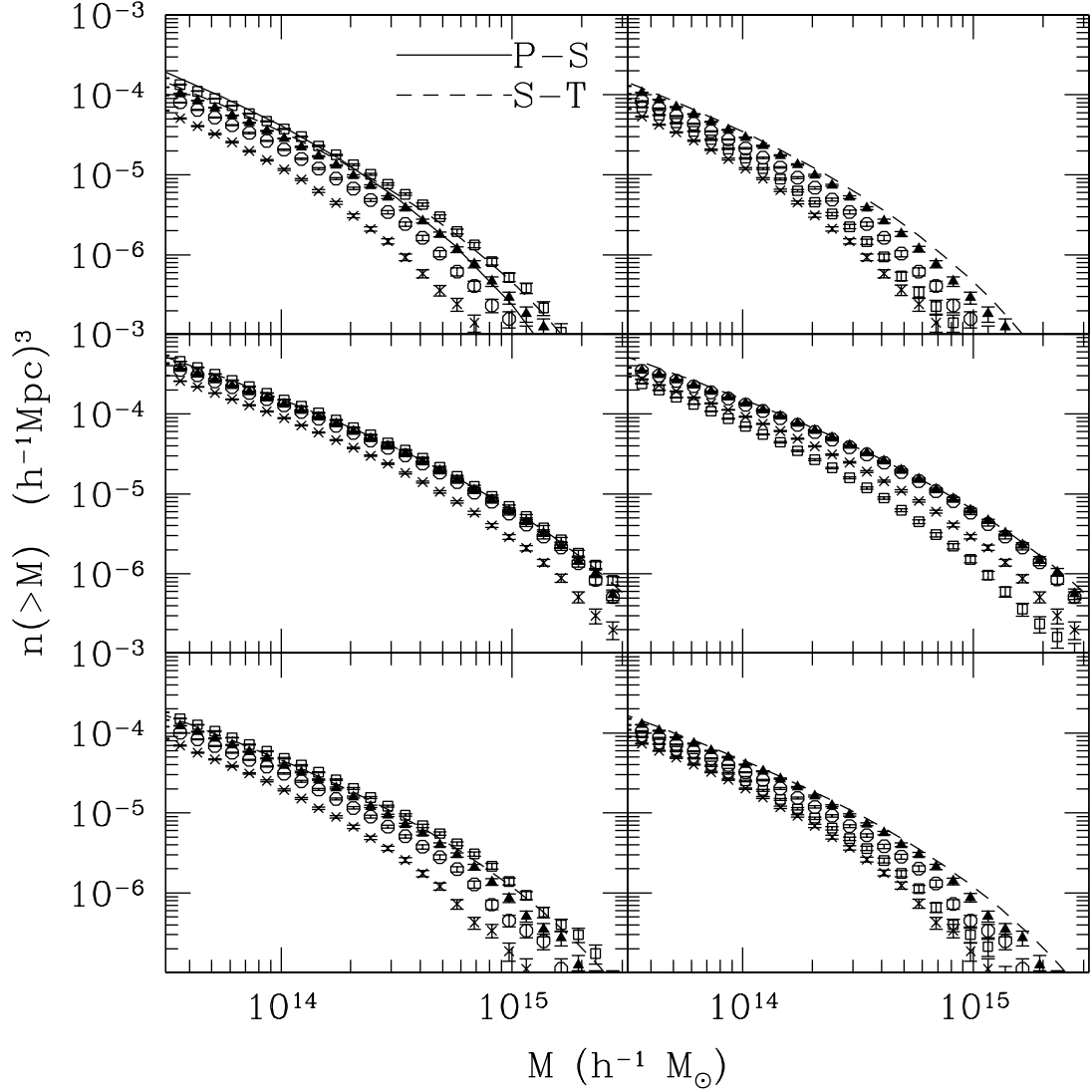}}
\end{center}
\caption{Mass functions for the 3 models, 2 linking lengths and 4 of the
mass definitions.  Left panels are for $b=0.2$, right for $b=0.1$.  Top to
bottom are Models 1, 2 and 3.  The open squares are FoF mass, solid
triangles $M_{\rm th-vir}$, open circles $M_{200c}$ and crosses $M_{500c}$.
In the first panel the solid line is the Press-Schechter prediction, and
the dashed line the fit to the Virgo simulations.  The solid line is
suppressed in the other panels.}
\label{fig:massfnC}
\end{figure*}

\begin{figure*}
\begin{center}
\resizebox{6in}{!}{\includegraphics{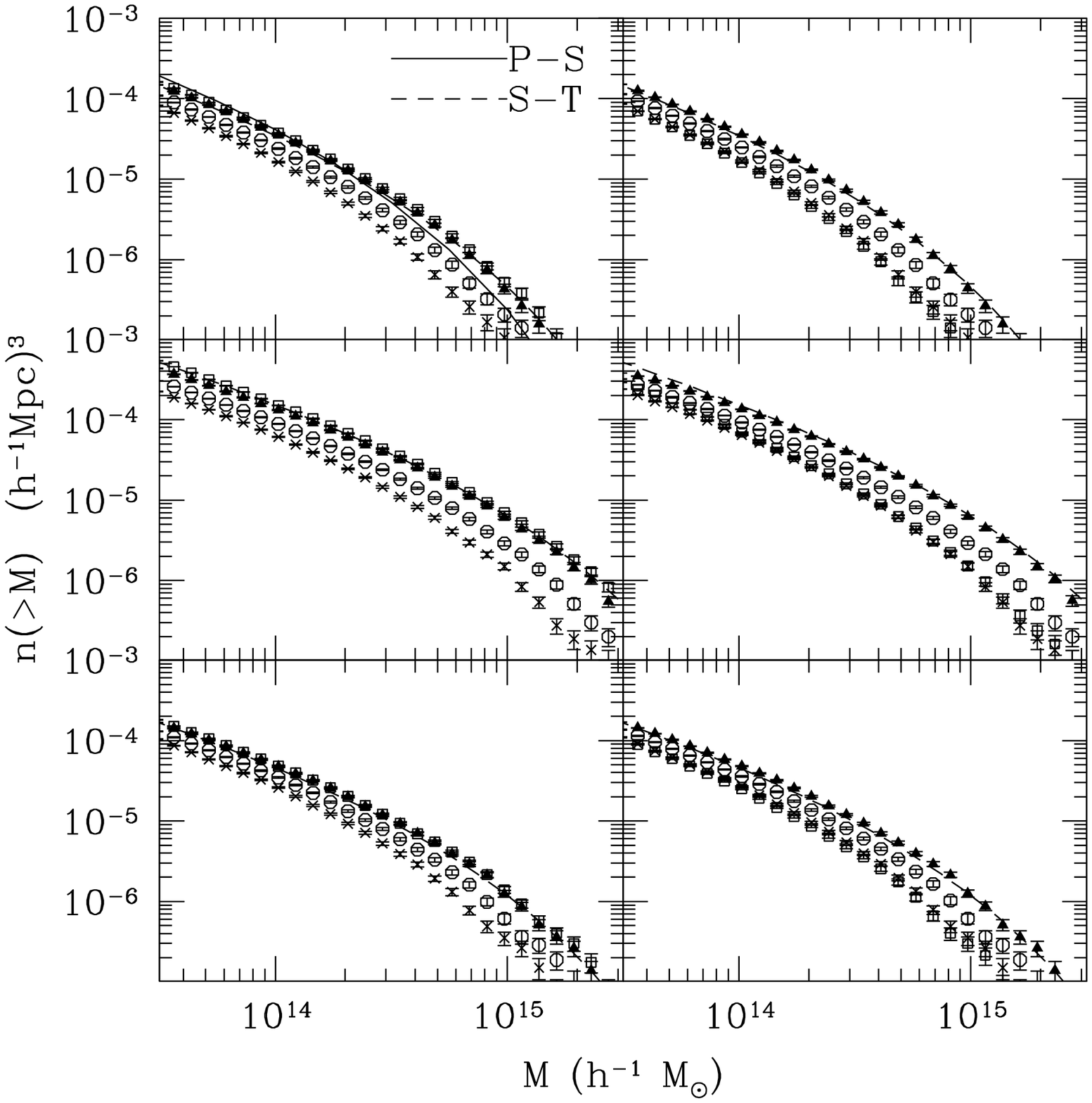}}
\end{center}
\caption{Mass functions for the 3 models, 2 linking lengths and 4 of the
mass definitions.  Left panels are for $b=0.2$, right for $b=0.1$.  Top to
bottom are Models 1, 2 and 3.  The open squares are FoF mass, solid
triangles $M_{180b}$, open circles $M_{500b}$ and crosses $M_{1000b}$.
In the first panel the solid line is the Press-Schechter prediction, and
the dashed line the fit to the Virgo simulations.  The solid line is
suppressed in the other panels.}
\label{fig:massfnB}
\end{figure*}

For completeness we would like to find a fit to the simulated mass functions
(Figs.~\ref{fig:massfnC}, \ref{fig:massfnB}).
Since to a very good approximation the mass function is independent of the
clustering of the halos, we can do this using the Poisson model.
First we bin the halos in mass using a large enough number of bins that no
bin contains more than 1 halo.  We use bins equally spaced in $\log M$.
Then we maximize the (log) likelihood
\begin{equation}
  -\log {\cal L} = \sum_{i\in{\rm full}} \log\mu_i - \sum_{j\in{\rm all}} \mu_j
  + {\rm const}
\end{equation}
where the sum on $i$ is over bins containing $1$ particle, the sum on $j$
is over all the bins and
\begin{equation}
  \mu \equiv {dn\over d\log M}\,\Delta\log M \ll 1
\end{equation}
is the mean number of halos per bin assuming a mass function $dn/d\log M$.
This method has the advantage of being independent of the chosen bin width
and correctly taking into account the Poisson statistics of the rare halos
at the high mass end.

We have chosen to use the modification to the Press-Schechter formula,
Eq.~(\ref{eqn:fnu}), proposed by Sheth \& Tormen~\cite{SheTor}, but to
allow $a$ and $p$ to be free parameters.
We maximize the likelihood for $a$ and $p$ fitting over the range
$M_0=10^{14}\,h^{-1}M_\odot$ to $M_1=3\times 10^{15}\,h^{-1}M_\odot$.
The results of this procedure are shown in Table~\ref{tab:fit}.
We found that the fit was somewhat sensitive to the particular value of
$M_0$ chosen, indicating that the numerical mass function wasn't perfectly
fit by the form of Eq.~(\ref{eqn:fnu}).  Using the higher resolution
simulation we found that the fitting function tended to overpredict the
abundance of halos with $\nu^2<1/2$, by a factor of almost 2 for the lowest
mass halos we could probe.  Since we concentrate here on the more massive
end of the mass function we did not attempt to correct for this.

There is a fair degree of variation in the best fit parameter $a$, while
$p$ is close to constant.  This is because $p$ essentially controls the
slope of the low-mass ($\nu\sim 1$) end of the mass function which is very
nearly the same for all the estimators.
We obtain reasonable agreement with Sheth \& Tormen \cite{SheTor} for the
conventional mass estimator $M_{200c}$ for the critical density cosmology
for example, but $a$ is significantly smaller using the $b=0.2$ FoF mass
and significantly larger using the $b=0.1$ FoF mass.
Generally $a$ increases as the mass estimator probes material primarily at
a higher density contrast.

For the `universal' contrast of $M_{180b}$ calculated directly from the
halo mass distribution we found better agreement with the numerical results
in all cases if we lowered $a$ slightly below the $0.7$ found by
Sheth \& Tormen for this mass estimator.
This is the opposite of the claim of Hu \& Kravtsov \cite{HuKra} that
$a$ should be increased to $0.75$ when using $M_{180b}$.
On the other hand for the one example we studied in detail where we converted
{}from a mass measured within $r_{500c}$ to one within $r_{180b}$ using an
NFW profile with $c=5$, the best fitting $a$ was slightly higher than $0.7$.
The difference in the mass functions so produced, as $a$ is changed from
$0.65$ to $0.75$, is at approximately the same level of the scatter from
model to model shown in Fig.~\ref{fig:unilin}.
Thus these fluctuations could be reflecting an intrinsic limit to how well
we can determine our fitting function.

\begin{table}
\begin{center}
\begin{tabular}{cccccc}
    &          & \multicolumn{2}{c}{$b=0.2$}&\multicolumn{2}{c}{$b=0.1$}\\
Sim & $\Delta$ & $a$   & $p$  & $a$  & $p$  \\ \hline
 1  &  FoF     & 0.64  & 0.34 & 1.17 & 0.31 \\
 1  & $\Dc$    & 0.79  & 0.32 & 0.79 & 0.32 \\
 1  &  200c    & 0.98  & 0.30 & 0.98 & 0.31 \\
 1  &  500c    & 1.36  & 0.29 & 1.35 & 0.30 \\
 1  &  180b    & 0.67  & 0.33 & 0.66 & 0.33 \\
 1  &  500b    & 0.89  & 0.31 & 0.89 & 0.31 \\
 1  & 1000b    & 1.13  & 0.29 & 1.12 & 0.30 \\ \hline
 2  &  FoF     & 0.64  & 0.32 & 1.41 & 0.27  \\
 2  & $\Dc$    & 0.66  & 0.29 & 0.65 & 0.29 \\
 2  &  200c    & 0.70  & 0.29 & 0.69 & 0.29 \\
 2  &  500c    & 1.07  & 0.26 & 1.05 & 0.27 \\
 2  &  180b    & 0.67  & 0.29 & 0.65 & 0.29 \\
 2  &  500b    & 1.07  & 0.26 & 1.05 & 0.27 \\
 2  & 1000b    & 1.48  & 0.25 & 1.45 & 0.26 \\ \hline
 3  &  FoF     & 0.64  & 0.34 & 1.20 & 0.30 \\
 3  & $\Dc$    & 0.76  & 0.31 & 0.76 & 0.31 \\
 3  &  200c    & 0.97  & 0.30 & 0.96 & 0.31 \\
 3  &  500c    & 1.38  & 0.28 & 1.37 & 0.29 \\
 3  &  180b    & 0.64  & 0.33 & 0.64 & 0.33 \\
 3  &  500b    & 0.87  & 0.30 & 0.86 & 0.31 \\
 3  & 1000b    & 1.13  & 0.29 & 1.12 & 0.30 \\
\end{tabular}
\end{center}
\caption{Mass function parameters for the different cosmologies and different
mass definitions (see text).}
\label{tab:fit}
\end{table}

\section{Clustering and the mass function} \label{sec:clustering}

Throughout we have assumed that the number of clusters in a given volume is
simply Poisson distributed about a mean value given by the mass function.
In principle however the positions of clusters are correlated and this can
induce non-Poisson fluctuations (Evrard et al.~\cite{EvrHubble}; for a
simple analytic model see Hu \& Kravtsov~\cite{HuKra}).
We expect this effect to be small when the objects are rare and the sample
region is large compared to the cluster correlation length
$r_0\sim{\cal O}(10\,h^{-1}{\rm Mpc})$ (see e.g.~Peebles \cite{LSSU})
but we can quantify its effect using numerical simulations.

Hu \& Kravtsov~\cite{HuKra} have investigated the additional scatter in the
normalization that arises from including clustering using a simple analytic
model where clusters are biased tracers of the linear density field.
We shall investigate this effect using the $z=0$ group catalog from the Hubble
Volume Simulation of a $\Lambda$CDM model run by the Virgo Supercomputing
Consortium (JFWCCECY).
To quantify the scatter in the mass function, we shall compute the best
fitting power spectrum normalization, $\sigma_8$, to 1000 random sub-volumes
of the simulation, using the maximum likelihood method described above.
We hold the other parameters fixed at their fiducial values for simplicity.
Each sub-volume is centered on a random point in the simulation, which can
be chosen as the center using the periodicity of the box.  Then all clusters
are kept which have $M$ above a threshold mass, are closer than $R$ to the
center of the box and would have $|b|>30^\circ$ if the box $x-y$ plane was
oriented parallel to the galactic plane.  This roughly mimics a {\it volume
limited\/} X-ray survey out to depth $R$ sensitive to the most massive, and
therefore most clustered but rarest galaxy clusters.
We choose two mass thresholds, $10^{14}\,h^{-1}\,M_\odot$ and
$3\times 10^{14}\,h^{-1}\,M_\odot$.
The scatter expected from a Poisson distribution can be estimated using the
same procedure, except that we first randomize the positions of the halos.

\begin{figure}
\begin{center}
\resizebox{3.5in}{!}{\includegraphics{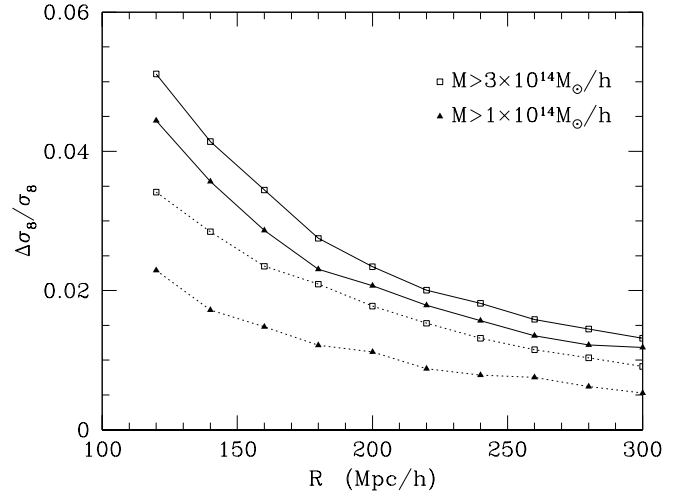}}
\end{center}
\caption{The standard deviation of $\sigma_8$, in units of the mean, as a
function of survey radius.  We plot a quantity, half the difference between
the $16^{\rm th}$ and $84^{\rm th}$ percentiles of the distribution, which
is slightly less sensitive to outliers than the variance; but the results
would be almost identical if we had plotted the variance.
Solid lines indicate the width of the distribution including the clustering
of clusters, dashed lines are for the randomized sample.  Lines joining
open squares are for $M>3\times 10^{14}\,h^{-1}\,M_\odot$ and joining solid
triangles for $M>10^{14}\,h^{-1}\,M_\odot$.}
\label{fig:variance}
\end{figure}

Fig.~\ref{fig:variance} shows the standard deviation in $\sigma_8$, in units
of the mean value, as a function of sampled volume for the `clustered' and
`Poisson' cases.  We checked that estimating the variance directly or from
the difference between the $16^{\rm th}$ and $84^{\rm th}$ percentiles of the
distribution gave similar results.
For all volumes studied the variance in $\sigma_8$ is increased by clustering
over the simple Poisson expectation (Evrard et al.~\cite{EvrHubble}), however
for volumes of interest ($R_{\rm max} > 300\,h^{-1}$Mpc) both the Poisson
variance and the increase due to clustering are almost negligible compared to
the other errors (see e.g.~Table 6 of Pierpaoli et al.~\cite{PSW}).
We checked explicitly that there was no bias in the mean introduced by the
neglect of clustering in the analysis.

\section{Conclusions} \label{sec:conclusions}

The multiplicity function, a measure of the number of halos per comoving
volume element per unit mass, is one of the central predictions of a model
of structure formation.
Dark matter dominated models in which structure evolves hierarchically from
gaussian initial conditions predict a mass function which is nearly
universal if expressed in the right units.  This is only true for a narrow
class of mass estimators, and is specifically not true for the estimators
which have been most commonly used up until now.

Given the complicated process by which halos form in hierarchical models,
the role of mergers and prevalence of sub-structure, it is highly convenient
that the mass function (in scaled units) is so close to universal.
We have provided fitting functions to the mass function from N-body
simulations for 8 different mass estimators, and shown how one can convert
between them.
We have found that measuring the mass of a halo using one definition and using
a simple average spherical profile, such as the NFW profile, to convert to
the `universal' $M_{180b}$ provides a remarkably good method of estimating
the mass function, even though individual halos show a large scatter among
different mass estimates.

Finally let us remark upon the small non-universality in the multiplicity
function, which can lead to misestimates of the true mass function if one
uses a fitting function like Eq.~(\ref{eqn:fnu}).
Neglecting the factor $d\log\sigma/d\log M$ in the mass function, making an
error of $\delta n/n$ in the number density per $\log M$ at mass $M$
translates into an error of $(\nu^2-1)^{-1} \delta n/n$ in the normalization
$\delta\sigma/\sigma$.
So for a typical cluster, with $\nu\sim 2-3$, the scatter in the mass function
doesn't limit our knowledge of $\sigma_8$ until the other uncertainties are
pushed below ${\cal O}(5\%)$.
Thus the non-universality of the mass function is not currently a limitation
to using the abundance of rich clusters to determine the normalization of the
matter power spectrum.
The uncertainties become increasingly important when it comes to using the
evolution of the mass function as a probe of $\Omega_{\rm m}$ or the equation
of state, $w$, of the dark energy.  For the latter, errors on $\sigma_8$
approaching the percent level are required.  For these ambitious measurements
it may not be sufficient to use a simple parameterized form for the
multiplicity function.  One could either resort to full blown numerical
simulations for a grid of models `near' the parameter region of interest or
attempt to find a `second variable' which correlates well with the scatter
between the simulation results and the P-S predictions.
As we approach the level of precision where these effects matter a variety of
other effects also become important, including the effects of clustering
(see e.g.~\S\ref{sec:clustering}, Evrard et al.~\cite{EvrHubble},
Hu \& Kravtsov~\cite{HuKra}) and how to treat merging systems.
It will be a challenge for theorists to keep the `theory uncertainty' below
the `experimental uncertainty' with the increasingly rapid advances in
observations.

\section*{Acknowledgments}

M.W.~would like to thank Marc Davis for comments on an early draft.
The simulations in this paper were carried out by the author at the
National Energy Research Scientific Computing Center and by the
Virgo Supercomputing Consortium using computers based at the Computing
Centre of the Max-Planck Society in Garching and at the Edinburgh parallel
Computing Centre.
The Virgo data are publicly available at http://www.mpa-garching.mpg.de/NumCos.
This work was supported in part by the Alfred P. Sloan Foundation.

\appendix

\section{The TreePM-SPH code}

The simulations in this paper were done using the {\sl TreePM-SPH\/} code
(White et al.~\cite{TreePM}) running in fully collisionless mode.  We
present a brief discussion of the features of the code here for completeness.

The {\sl TreePM\/} code was specifically designed to run on distributed memory
computers or clusters of networked workstations and evolves dark matter (and
gas) in a periodic simulation volume.
The code uses the {\sl TreePM\/} method of Bagla~\cite{Bagla} for the
gravitational force and smoothed particle hydrodynamics
(SPH; Lucy~\cite{Luc} and Gingold \& Monaghan~\cite{GinMon}) in its
`entropy' formulation (Springel \& Hernquist~\cite{SprHer}) to compute
the hydrodynamic forces.  The (collisionless) dark matter component and
(collisional) gas are assumed to interact only through gravity.
The code is written in standard C and uses the Message Passing Interface (MPI)
communications package, making it easily portable to a variety of parallel
computing platforms.  It performs dynamical load balancing and scales
efficiently with increasing numbers of processors.  It will run on an
arbitrary number of processors, though it is slightly more efficient if the
number is a power of two.

The particles are integrated using a second order leap-frog method, where
the relevant positions, energies etc are predicted at a half time step and
used to calculate the accelerations which modify the velocities.
The time step is dynamically chosen as a small fraction (depending on the
smoothing length) of the local free-fall time.  Particles have individual
time steps so that the code can handle a wide range in densities efficiently.
To increase speed the force on any given particle is computed in two stages.
The long-range component of the force is computed using the PM method,
while the short range component is computed from a {\it global\/} tree.
In this manner the code is similar in spirit to P${}^3$M except that the short
range force, being computed from a tree, scales as $N\log N$ rather than $N^2$.

Rather than a Plummer potential we use a spline softened force.  We use the
same smoothing kernel for the gravity and SPH calculations.  The long-range
force is smoothed on 2 grid cells, and the opening criterion for the tree is
set to achieve $1\%$ accuracy in the short-range force.  With these standard
parameters the 90th percentile force error is 1.2\% for lightly clustered
distributions while for very uniform distributions the 90th percentile error
rises slightly to 1.9\%, as the total force is smaller and the short range
force contributes less.

We have made extensive comparisons of the code described here to other codes
described in the literature, building on the many test problems that those
codes have been shown to satisfy.  In particular we have compared
{\sl TreePM\/} extensively with {\sl Gadget\/} (Springel, Yoshida \&
White~\cite{SprYosWhi}).  Details of these comparisons, along with results
from self-similar evolution, hydrodynamics tests including the sod shock
tube, the Santa-Barabara cluster comparison project
(Frenk et al.~\cite{Fre})
etc can be found in White, Springel \& Hernquist~\cite{TreePM}.


\begin{thebibliography}{99}
\bibitem[1999]{Bagla}
  Bagla J., 1999, preprint [astro-ph/9911025]
\bibitem[1991]{BCEK}
  Bond J.R., Cole S., Efstathiou G., Kaiser N., 1991, \apj, 379, 440
\bibitem[1996]{BonMye}
  Bond J.R., Myers S., 1996, \apjs, 103, 41
\bibitem[1991]{Bow}
  Bower R.J., 1991, \mnras, 248, 332
\bibitem[1989]{ColKai}
  Cole S., Kaiser N., 1989, \mnras, 237, 1127
\bibitem[1985]{DEFW}
  Davis M., Efstathiou G., Frenk C.S., White S.D.M., 1985, \apj, 292, 371
\bibitem[1988]{EFWD}
  Efstathiou G., Frenk C.S., White S.D.M., Davis M., 1988, \mnras, 235, 715
\bibitem[1988]{EfsRee}
  Efstathiou G., Rees M., 1988, \mnras, 230, 5P
\bibitem[2002]{EvrHubble}
  Evrard A.E., et al., 2002, \apj, in press [astro-ph/0110246]
\bibitem[1999]{Fre}
  Frenk C.S., et al., 1999, \apj, 525, 554
\bibitem[1994]{GelBer}
  Gelb J., Bertschinger E., 1994, \apj, 436, 467
\bibitem[1977]{GinMon}
  Gingold R.A., Monaghan J.J., 1977, \mnras, 181, 375
\bibitem[2002]{HuKra}
  Hu W., Kravtsov A., 2002, preprint [astro-ph/0203169]
\bibitem[1998]{Jen98}
  Jenkins A., Frenk C.S.,  Pearce F.R., Thomas P.A., Colberg J.M.,
  White S.D.M., Couchman H.M.P., Peacock J.A., Efstathiou G., Nelson A.H.,
  1998, \apj, 499, 20
\bibitem[2000]{JFWCCECY}
  Jenkins A., Frenk C.S., White S.D.M., Colberg J.M., Cole S., Evrard A.E.,
  Couchman H.M.P., Yoshida N., 2001, MNRAS, 321, 372 (JFWCCECY)
\bibitem[1993]{LacCol1}
  Lacey C., Cole S., 1993, \mnras, 262, 627
\bibitem[1994]{LacCol2}
  Lacey C., Cole S., 1994, \mnras, 271, 676
\bibitem[1977]{Luc}
  Lucy L., 1977, \aj, 82, 1013
\bibitem[1996]{MoWhi}
  Mo H.J., White S.D.M., 1996, \mnras, 282, 347
\bibitem[1996]{NFW}
  Navarro J., Frenk C.S., White S.D.M., 1996, \apj, 462, 563
\bibitem[1999]{Peacock}
  Peacock J.A., 1999, Cosmological Physics, Cambridge University Press,
  Cambridge
\bibitem[1996]{PD96}
  Peacock J.A., Dodds S.J., 1996, \mnras, 280, L19
\bibitem[1990]{PeaHea}
  Peacock J.A., Heavens A., 1990, \mnras, 243, 133
\bibitem[1980]{LSSU}
  Peebles, P.J.E., 1980, The Large-Scale Structure of the Universe,
    Princeton University Press, Princeton, \S 36.
\bibitem[1993]{PPC}
  Peebles, P.J.E., 1993, Principles of Physical Cosmology, Princeton University
  Press, Princeton, chapter~25.
\bibitem[2001]{PSW}
  Pierpaoli E., Scott D., White M., 2001, \mnras, 325, 77
\bibitem[1974]{PS}
  Press W.H., Schechter P., 1974, \apj, 187, 452
\bibitem[2002]{Sel}
  Seljak U., 2002, preprint [astro-ph/0111362]
\bibitem[1999]{SheTor}
  Sheth R., Tormen G., 1999, \mnras, 308, 119
\bibitem[2002]{SprHer}
  Springel V., Hernquist L., 2002, \mnras, in press [astro-ph/0111016]
\bibitem[2001]{SprYosWhi}
  Springel V., Yoshida N., White S.D.M., 2001, New Astronomy 6, 79
  [astro-ph/0003162]
\bibitem[2001]{HaloMass}
  White M., 2001, A\&A, 367, 27 [astro-ph/0011495]
\bibitem[2002]{TreePM}
  White M., Springel V., Hernquist L., 2002, in preparation.
\bibitem[1993]{WEF}
  White S.D.M., Efstathiou G., Frenk C., 1993, \mnras, 262, 1023
\bibitem[2002]{ZTWB}
  Zheng Z., Tinker J.L., Weinberg D.H., Berlind A.A., 2002, preprint
  [astro-ph/0202358]
\end{thebibliography}
\end{document}